\documentclass[12pt,a4paper]{article}
\usepackage{jheppub}

\usepackage{graphicx, epsfig}
\usepackage{color}

\usepackage{amsmath}
\usepackage{amssymb}
\usepackage{amsthm}
\newtheorem{thm}{Theorem}
\begin{document}


\vskip 0.25in

\newcommand{\todo}[1]{{\bf ?????!!!! #1 ?????!!!!}\marginpar{$\Longleftarrow$}}
\newcommand{\nn}{\nonumber}
\newcommand{\tr}{\mathop{\rm Tr}}
\newcommand{\gen}[1]{\langle #1 \rangle}
\global\long\def\comment#1{}

\begin{center}
{\LARGE \bf Exploring the Potential Energy Landscape Over a Large Parameter-Space}
\end{center}

\medskip

\vspace{.4cm}

\begin{center}
{\large Yang-Hui He}$^1$,
{\large Dhagash Mehta}$^2$,
{\large Matthew Niemerg}$^3$,\\
{\large Markus Rummel}$^4$ \&
{\large Alexandru Valeanu}$^5$
\end{center}

\vspace*{1.0ex}

\begin{center}
{\it
{\small
{${}^{1}$
Department of Mathematics, City University, London, EC1V 0HB, UK;\\
School of Physics, NanKai University, Tianjin, 300071, P.R.~China;\\
Merton College, University of Oxford, OX14JD, UK\\
\qquad hey@maths.ox.ac.uk\\
}
\vspace*{1.5ex}
{${}^{2}$
Department of Physics, Syracuse University, Syracuse, NY 13244, USA.\\
\qquad dbmehta@syr.edu \\
}

\vspace*{1.5ex}
{${}^{3}$
Department of Mathematics, Colorado State University, Fort Collins, CO 80523, USA. \\
\qquad niemerg@math.colostate.edu
}

\vspace*{1.5ex}
{${}^{4}$
II. Institut f\"ur Theoretische Physik der Universit\"at Hamburg, D-22761 Hamburg, Germany\\
\qquad markus.rummel@desy.de
}

\vspace*{1.5ex}
{${}^{5}$
Trinity College, University of Oxford, OX14JD, UK\\
\qquad alexandru.valeanu@trinity.ox.ac.uk
}
}}
\end{center}

\vspace*{4.0ex}
\centerline{\textbf{Abstract}}
\noindent

\comment{
Finding all the stationary points of a realistic potential energy landscape (PEL) is a difficult problem as it amounts to solving non-linear equations efficiently. Finding them over a large parameter-space is even more difficult problem as one has to solve the equations individually for each parameter-point. Essentially, the task boils down to solving a parametric system of equations over a large parameter space. We introduce two algebraic geometry based methods which efficiently solve such systems: a symbolic method called the Comprehensive Gr\"obner basis method, and a numerical method called the cheater's homotopy. We then specialize to a few potentials arising from string phenomenology and find the stationary points that are minima, called string vacua, over a large parameter space. Even with our modest computational resources we have been able to scan over just under $5$ hundred thousand parameter-points for a well-known string theory compactification manifold called the quintic. We also extract several
important physical quantities such as the gravitino and moduli masses and the string coupling.
}

Solving large polynomial systems with coefficient parameters are ubiquitous and constitute an important class of problems.
We demonstrate the computational power of two methods--a symbolic one called the Comprehensive Gr\"obner basis and a numerical one called the cheater's homotopy--applied to studying both potential energy landscapes and a variety of questions arising from geometry and phenomenology.
Particular attention is paid to an example in flux compactification where important physical quantities such as the gravitino and moduli masses and the string coupling can be efficiently extracted.

\newpage

\tableofcontents

\vspace{0.5in}


\section{Introduction}
The hypersurface defined by a potential energy function $V(\vec{x}; \vec{a})$, with $\vec{x} = (x_1, \dots, x_N)$ being the variables and $\vec{a} = (a_1, \dots, a_k)$ being a set of parameters, is called the potential energy landscape (PEL) of a given physical model.
The landscape, in particular, refers to the potentially large number of parameters which constitutes a moduli space.
The special points of a PEL, defined as the critical (or stationary) points of $\frac{\partial V(\vec{x})}{\partial x_{i}}=0,$ with $1 \leq i \leq N$, give crucial information about the physical system depending on the problem at hand.

In recent years, a huge influx of advancements have come through from many areas in theoretical physics and chemistry in understanding the PEL and its relation to various physical and chemical properties. The research areas where the PEL methods have have been immensely successful in explaining the underlying physics or chemistry include clusters \cite{Wales:04,Wales05,Wales06,WalesB06,MiddletonW03,walesd01}, disordered systems and glasses~\cite{binder1986spin,RevModPhys.80.167}, biomolecules, protein folding, string phenomenology~\cite{Ibanez:2012zz}, and within this area flux compactifications~\cite{Dasgupta:1999ss,McOrist:2012yc,Giddings:2001yu,Douglas:2006es,Marsh:2011aa,Grana:2005jc,Blumenhagen:2006ci}.

Due to the importance of these critical points of the PELs, finding them in realistic models has been an active research area for quite some time.
The stationary equations for any realistic model are invariably non-linear which are known to be extremely difficult to solve in general.
Various numerical techniques exist based on the Newton-Raphson method and its sophisticated variants~\cite{wales92,wales93d,broderix2000energy,doye2001saddle,wales2003stationary} where a random initial guess is refined to attain a single solution of the system.
However, in all these methods, even after feeding a large number of random initial guesses, one can never be sure of obtaining all the solutions.

While \textit{many} solutions as opposed to \textit{all} solutions may be fine in many applications, often one has to find all the solutions.
The problem of not obtaining all the solutions becomes crucial when one needs the information about all the local minima or the global minimum, etc.
For example, in string theory models, where in many cases the potential energy landscape is defined by an effective four-dimensional, $\mathcal{N}=1$ supergravity scalar potential $V(K,W)$ given a K\"ahler potential $K$, and a superpotential $W$, one has to find all the minima. These minima are called {\em string vacua} and addressing the plethora of solutions is one of the most important of current theoretical challenges.

Fortunately, in a large class of models, the equations are polynomials, or at least they have polynomial-like non-linearity.\footnote{
In the presence of non-perturbative effects where exponential terms contribute, one could still reduce the system into polynomials of Lambert functions, perform standard polynomials manipulations and treat the Lambert functions numerically \cite{Gray:2007yq}.
}
Once they are identified as a system of polynomial equations, we can use techniques from algebraic geometry, or more specifically, computational algebraic geometry to solve them by systematically transforming a given system of multivariate polynomial equations to another system which is \textit{easier} to solve and has the same solutions as the original one. The new system of equations is called a {\em Gr\"obner basis} (GB), and the algorithm to compute one is called the Buchberger algorithm (BA).

The biggest advantage here is that one can find all the solutions of the given system once the computation is finished.
Computational algebraic geometry, which is essentially a set of techniques based on the Gr\"obner basis method, has become one of the most useful tools to study a number of phenomena in theoretical physics. Recently, the rich interplay between algebraic geometry and theoretical physics, especially in gauge and string theory, has been an active area of research \cite{comp-book}. Activities in these areas have been enhanced with the increased power of computers and the development of algorithms in computational algebraic geometry.

More specifically, a variety of methods have been used to study the moduli space of vacua over the past few years \cite{Gray:2006jb,Gray:2008yu} based on symbolic computational algebraic geometry, most of whose sub-methods and sub-algorithms rely on Gr\"obner basis techniques (cf.~\cite{Gray:2009fy} for an overview on the method).
For convenience, a freely available computational package, {\sf StringVacua}, which is a {\sf Mathematica} package specifically designed for phenomenologists \cite{Gray:2006gn,Gray:2008zs,Gray:2007yq} exists.  {\sf StringVacua} interfaces with the advanced computational algebraic geometry package {\sf Singular} \cite{DGPS}.
Using {\sf StringVacua}, one can extract important information such as the dimension of the vacuum, the number of real roots in the system, stability and supersymmetry of the potential, or the branches of moduli space of vacua, etc., using a regular desktop machine in many circumstances.

However, the GB method is known to suffer from {\it exponential complexity}, i.e., the computation time and the RAM required by the BA algorithm increases exponentially with the number of variables, equations, degree, and terms in each polynomial; it is usually less efficient for systems with irrational coefficient parameters $\vec{a}$; and it is also highly sequential, i.e., very difficult to parallelize the algorithm and put on a big cluster.

To overcome these shortcomings of the GB methods, a different approach called numerical algebraic geometry (NAG) was recently introduced.
Its core algorithm, called {\em numerical polynomial homotopy continuation} (NPHC), guarantees to find all of the solutions unlike other numerical methods such as the Newton-Raphson and its variants.
Moreover, unlike the Gr\"obner basis techniques, the NPHC method is ``embarrassingly parallelizable'', and hence one can solve more complicated systems efficiently using computer clusters.

NAG was introduced in particle theory and statistical mechanics areas in Ref.~\cite{Mehta:2009}.
Subsequently, the NPHC method was used to solve systems arising in numerous physical phenomena in lattice field theories \cite{Mehta:2009zv,Hughes:2012hg},
statistical physics \cite{Mehta:2011xs,Kastner:2011zz,Mehta:2012qr,Nerattini:2012pi}, particle phenomenology \cite{Maniatis:2012ex,CamargoMolina:2012hv}, and string phenomenology \cite{Mehta:2011wj,Mehta:2012wk,Hauenstein:2012xs}.

\subsection{Parametric Potentials}
As mentioned above, in generic physical applications, the potential energy function is defined over a possibly vast parameter-space, where each point $\vec{a}$ represents a different physical situation.
For example, in a statistical mechanics model, the parameters are the disorders \cite{binder1986spin}.  Another example in theoretical chemistry where the models are Morse clusters, the parameters represent the strength of the inter-particle potential \cite{Wales01,Wales10b,CalvoDW12,Wales12}. As a third example, in lattice field theory the parameters represent different background fields in the models~\cite{Mehta:2009zv,Hughes:2012hg,vonSmekal:2007ns,vonSmekal:2008es}. Another situation is in flux compactifications within string phenomenology where the parameters represent the flux quanta.

The flux quanta are discrete parameters that are given as integrals of $n$-forms along $n$-cycles in a compact space, see \S\ref{examples_sec}.
Their discreteness arises as a Dirac quantization condition. For many compactification manifolds, for instance Calabi-Yau spaces, these cycles exist on the order of 10 to 100  where the flux quanta can be chosen independently.
This yields an exponentially large parameter space that is only limited by conditions of conservation of certain charges in the compact space.

The parametric potentials adds one more, rather severe, problem where as the parameters enter the system of equations, one has to solve the system of parametric equations.
This is in fact a quite difficult problem.  If we use the above mentioned methods directly, we have to specify numerical values of the parameters \textit{before} using the methods.
In other words, we have to solve the system separately from scratch for each point in the parameter space. In practice, due to the large number of physically interesting parameter-points, this crude way of solving the problem becomes prohibitively time consuming as well as computationally expensive.
Traditionally, one resorts to assigning generic, random values to these parameters. However, this misses the intricacies of special parameters.

In this paper, we introduce two methods which deal with parametric systems extremely efficiently.
The first method is called the {\bf Comprehensive Gr\"obner basis} (CGB) method. This is a completely symbolic method where given a parametric system of equations, CGB yields a new system, leaving all the parameters in the symbolic form, which is a GB for all values of the parameters and also for all the \textit{specializations}, i.e., special values of the parameters. Once we obtain a CGB for a given system, it then only amounts to inputting the specific values of parameters and extracting the corresponding solutions out.
Thus, we can efficiently solve the system for as many parameter-points as we desire since the system is \textit{solved} for the whole parameter space. However, as one may surmise, this method has the same aforementioned short-comings as the usual GB method.
Nevertheless, when successful in finishing the computation, the CGB method reduces the amount of work in finding string vacua over a large space of flux parameters drastically.

The second method is called the {\bf cheater's homotopy} (CH) method, which is based on NAG.
In NPHC, one first has to estimate an upper bound on the number of solutions of the given system so that one can then construct another system which has the same number of solutions as the estimate.
To reduce the computation in this method, we must come up with a tighter upper bound on the number of solutions.  Three of the most important solution bounds were demonstrated in our previous works \cite{Mehta:2009,Mehta:2009zv,Mehta:2011xs,Mehta:2012wk}.
However, in most of the realistic systems, the sparsity of the systems (i.e., the number of monomials in each equation may be only a few) is not fully taken into account.
Thus, even though the original system may have only a few solutions, we may have to track many paths making it computationally expensive. Cheater's homotopy relies on the fact that the maximum number of solutions of a parametric system of polynomial equations over all the parameter-points is that at a generic parameter-point. The number of paths to be tracked is usually drastically smaller than any of the other upper bounds for the cheater's homotopy \cite{li1989cheater}.

The second, and perhaps the most crucial, advantage is that in the cheater's homotopy we obtain the start system once for for a generic parametric system, i.e., we then can use the same start system for arbitrary parameter-points in the entire parameter-space.
Several more advantages of this method exist when solving these systems on a computer cluster, but most importantly the method is ``embarrassingly parallelizable'', a subject matter to which we shall later return.
A very efficient package, which is yet to be publicly available, called {\sf Paramotopy} \cite{paramotopytoappear}, uses all the computational advantages of the method in our favor and can deal with hundreds of thousands of parameter-points. In this paper, we heavily rely on {\sf Paramotopy} for our computations.

We would like to make the clarification that the above two methods are based on complex algebraic geometry which means that the variables are first considered to be complex, and then once the solutions are obtained, only purely real solutions are retained for analysis if the system was supposed to be made of real variables.
However, there is a more recent method from real algebraic geometry, based on the discriminant varieties~\cite{hanan2010stability,hernandez2011towards}, which treats both variables and parameters as reals from the beginning and gives a different set of information than the methods described in this paper.
The details on this method with applications arising from physics will be addressed in forthcoming works.

We first introduce the Comprehensive Gr\"obner basis method and solve a few toy models in \S\ref{CGB_sec}. Then, to make the paper self-contained, we briefly explain the NPHC method before explaining the cheater's homotopy method in \S\ref{paramotopy_sec}. Therein, we also introduce the salient features of the {\sf Paramotopy} package.
In \S\ref{examples_sec}, we solve several toy models as well as more realistic models such as flux compactifications on the quintic manifold. We will extract some interesting physics using the string vacua of these models over a large space of fluxes. Finally, we conclude with remarks and prospects in \S\ref{conc_sec}.

\section{Algebraic Geometry \& the Comprehensive Gr\"obner Basis Method}\label{CGB_sec}
In this section, we explain the concept of a Comprehensive Gr\"obner basis in more detail and some basics of algebraic geometry to set the nomenclature and also to facilitate the reader.
We first introduce few technical terms leading to the definition of a Gr\"obner basis. Then, we will explain what a Comprehensive Gr\"obner basis is.
Readers uninterested in the technicalities involved in the definition may freely skip subsection \S\ref{s:CGB} after reading the next two paragraphs.

Roughly speaking, given a system of polynomial equations,
the Buchberger Algorithm (BA), or its refined variants,
compute a new equivalent system of polynomials, called a Gr\"obner basis \cite{CLO:07} which has nicer properties; using the BA on multivariate polynomial systems is analogous to Gaussian elimination for linear systems.
Nowadays, efficient variants of the BA are available, e.g., F4 \cite{Faugere99anew}, F5 \cite{Faug:02}, and Involution Algorithms~\cite{2005math1111G}.
Symbolic computation packages such as {\sf Mathematica}, {\sf Maple}, {\sf Reduce}, etc., have built-in commands to calculate a Gr\"obner basis.
Moreover, {\sf Singular} \cite{DGPS}, {\sf COCOA} \cite{CocoaSystem}, and {\sf Macaulay2} \cite{M2}
are specialized packages for computational algebraic geometry available freely.  {\sf MAGMA} \cite{BCP:97} is also such a specialized package available commercially.

A system of polynomial equations with parameters is called a parametric system; finding the critical points of a polynomial PEL is precisely such a system.
If one is interested in solving the system at finitely many points on the parameter-space, then inserting the numerical values of the parameters in the system and obtaining a Gr\"obner basis is a quick escape, especially for a small number of parameters.
A natural question to ask is if it is possible to obtain a Gr\"obner basis for a given monomial ordering in terms of the symbolic form of the parametric coefficients, valid for all its special cases, called \textit{specializations}. One can indeed compute such a `parametric Gr\"obner basis' called Comprehensive Gr\"obner basis (CGB) \cite{weispfenning1992comprehensive}.
Algorithmically, we use the internal libraries of {\sf Singular} to compute the CGB in this paper.

\subsection{The Comprehensive Gr\"obner Basis}\label{s:CGB}
The technicalities in this subsection will lead to a very useful result, namely that we can transform a given system of multivariate polynomial equations to another one which has the same solutions but is easier to solve.
Here, the original system is considered as a basis of an algebraic object, called an \textit{ideal}.
Then an important result, that an appropriate change of this basis leaves the solution space unchanged, is used.

\subsection*{Polynomial Rings}
We define a polynomial $f$ as
\begin{equation}
f = \sum_{\alpha} a_{\alpha}x^{\alpha}
.
\end{equation}
Here, the sum is over a finite number of $m$-tuples $\alpha = (\alpha_{1},\dots ,\alpha_{m})$ and $x^{\alpha} = x_{1}^{\alpha_{1}}\dots x_{m}^{\alpha_{m}}$ is a monomial with all $\alpha_{i}$ being non-negative integers.
The coefficients $a_{\alpha}$ and the variables $x_i$ take values from the field $K$.
However, to use the results of Algebraic Geometry to its full extent, unless otherwise specified, we will take $K = \mathbb{C}$.

Now, if $K[x_{1},\dots ,x_{m}]$ is the set of all polynomials in variables
 $x_{1},\dots ,x_{m}$ with coefficients in $K$, then $f \in K[x_{1},\dots ,x_{m}]$ can be viewed as a function $f:K^{m}\longrightarrow K$ where $K^{m}$ is the affine space of all coefficients.
Thus, the sum and product of two polynomials is a polynomial, and a polynomial $f$ divides a polynomial $g$ if and only if $g = fh$, for some $h\in K[x_{1},\dots ,x_{m}]$. Using this, it can be shown that under addition and multiplication, $K[x_{1},\dots ,x_{m}]$ satisfies all of the field axioms except for the existence of a multiplicative inverse because $\frac{1}{x}$ is not a polynomial.
Indeed, $K[x_{1},\dots ,x_{m}]$ satisfies the axioms for a commutative ring, or more precisely a {\em polynomial ring} \footnote{For a nice discussion on related topics, the reader is referred to~\cite{CLO:07, CLO:98}.}.

\subsection*{Ideal}
One can now view all the polynomials of a system of polynomial equations as \textit{elements} of a polynomial ring.
Hence, one can also define a corresponding \textit{vector space}, called an ideal. More specifically, an ideal $I$ is a subset of $K[x_{1},\dots ,x_{m}]$ with the following properties:

\begin{enumerate}
\item $0\in I$,
\item $f+g \in I$ for all $f,g \in I$, and
\item $hf \in I$ for $f \in I$ and $h \in K[x_{1}\dots, x_{m}]$.
 \end{enumerate}

Consider any $f \in I \subset K[x_{1},\dots ,x_{m}]$.  If $f$ can be written as $f = \sum_{\alpha} a_\alpha h_\alpha$ with $a_\alpha \in K$ and $h_{\alpha} \in K[x_{1},\dots, x_{m}]$, then we write $I\; =\; \gen{h_{\alpha}} \subset K[x_{1},\dots ,x_{m}]$.  If the indexing set $\alpha$ is finite, say with cardinality $t$, then $I$ is called a \textit{finitely generated ideal}. The polynomials $h_{1},\dots ,h_{t}$ are then said to be a \textit{finite basis} of $I$, and we write $I = \gen{h_1, \dots, h_t}$.

\subsection*{Affine Variety}
So far, we have introduced the algebraic counterpart of Algebraic Geometry. The solution space of a given ideal is called a \textit{variety}. Specifically, an affine variety of an ideal $I = \; \gen{h_{1},\dots ,h_{t}}$ is the set of common zeros of polynomials $h_{1}\dots ,h_{t}$ in affine space, denoted as $V(h_{1},\dots ,h_{t})$ or $V(I)$.

\subsection*{Gr\"obner Basis}
The formalism of Algebraic Geometry turns out to be very helpful.
Interpreting the polynomials $h_{i}$ as a basis of $I$, we can \textit{change the basis} to, say, $\gen{H_{1},\dots ,H_{s}}$. Then, it can be shown that the solution space remains unchanged in an appropriate change of basis, that is, $V(h_{1},\dots ,h_{t}) = V(H_{1},\dots ,H_{s})$.
In essence, we use computational techniques to find a basis that is easier to deal with than the original one, in a certain sense. Such a basis is called a Gr\"obner basis.

In linear algebra, such a change of basis can be done via Gaussian Elimination, and the new basis is the familiar Row-Echelon form.
In general, an algorithm to obtain a Gr\"obner basis performs a specific set of algebraic operations including factorizing and dividing the polynomials.
In any algorithm that computes a Gr\"obner basis, the division requires one to impose a \textit{total order} among the monomials.
This is called a \textit{monomial ordering}.

A monomial ordering is a relation `$\succ$'
on the set of monomials $x^{\alpha}$, $\alpha \in \mathbb{Z}_{\ge 0}^{n}$, satisfying the following properties:

\begin{enumerate}
\item the ordering always tells which of two distinct monomials is greater,
\item the relative order of two monomials does not change when they are each multiplied by the same monomial, and
\item every strictly decreasing sequence of monomials eventually terminates~\cite{SW:95}.
\end{enumerate}

Different types of monomial orderings exist that satisfy the aforementioned properties such as lexicographic, graded lexicographic, graded reverse lexicographic, or degree lexicographic. Different monomial orderings are useful depending on the algorithm that is employed to compute the Gr\"obner Basis.
Lexicographic orderings will be primarily used throughout our discussion. To learn more about monomial ordering, the reader is referred to~\cite{CLO:07,CLO:98}.

By fixing a monomial order, we define a \textit{leading term} for each
 polynomial of a given ideal, denoted as $\gen{LT(h_{1}), \dots ,LT(h_{t})}$.
One can always find a finite subset $G = \; \gen{ H_{1},\dots ,H_{s} }$
 of an ideal $I$ (except for the trivial case $I = \gen{0}$) such that every leading term of $f\in I$ can be generated by
 $\gen{LT(H_{1}),\dots ,LT(H_{s})}$.
Here, $f\in I$ means that $f$ is an algebraic combination of $h_{1},\dots ,h_{t}$, as is required for $I$ to be an ideal.
Such a subset $G$ is called a Gr\"obner basis with respect to the specific monomial order.\footnote{It should be noted here that a Gr\"obner basis may not be unique for a fixed monomial ordering. So, we call it \textit{a} Gr\"obner basis rather than \textit{the} Gr\"obner basis. However, the so-called reduced Gr\"obner basis is unique for a given monomial ordering. The reader is referred to Ref.~\cite{CLO:07, CLO:98} for more details.}

One can show that for any given monomial order, every nontrivial ideal $I\subset K[x_{1},\dots ,x_{m}]$, has a Gr\"obner basis and that any Gr\"obner basis for an ideal $I$ is a basis of $I$. One can also show that $V(I)$ can be computed by any basis of $I$, and so the solutions of $I$ are the same as that of any of its Gr\"obner basis for any monomial ordering.

A well-defined procedure exists to compute a Gr\"obner basis for any given ideal and monomial ordering, called the Buchberger algorithm. It should be noted that the Buchberger algorithm reduces to Gaussian elimination in the case of linear equations, as it is a generalization of the latter.
Similarly, it is a generalization of the Euclidean algorithm for the computation of the Greatest Common Divisors of a univariate polynomial.

\subsection*{Comprehensive Gr\"obner Basis}
If the leading coefficient of each element of the basis is $1$ and no monomial in any element of the basis is in the ideal generated by the leading terms of the other elements of the basis, the basis is called a reduced Gr\"obner basis.  A reduced Gr\"obner basis is unique for a given ideal and monomial ordering, unlike a Gr\"obner basis.

Now, if we have a parametric ideal, i.e., $I= \gen{f_1, \dots, f_s} \subset R[x_1,\dots, x_n; a_1,\dots, a_m]$, where $R$ is a unique factorization domain, then a Comprehensive Gr\"obner basis (CGB) is the distinct reduced Gr\"obner basis for all possible values of the parameters $a_1,\dots, a_m$ \cite{weispfenning1992comprehensive}. There are several algorithms available to compute the CGB \cite{montes2002new,suzuki2003alternative,kapur2012efficient}. We refer the reader willing to learn more about the actual algorithm and related issues to these references, and leave the section by noting that we use the internal libraries of {\sf Singular} to compute the CGB in this paper.

Let us illustrate with a trivial example of a nonlinear parametric equation $a x^2 + b x + c=0$; this equation defines an ideal in $\mathbb{C}[x]$ with parameters $a,b,c$ in one variable $x$.
Singular's CGB library yields that there are three cases:
\begin{enumerate}
\item Case-1: for $a=b=c=0$, the solution is the whole $\mathbb{C}$, i.e. all $x \in \mathbb{C}$;
\item Case-2: for $a=0$ and $b\neq 0$, the solution is the line $b x + c=0$; and
\item Case-3: for $a\neq 0$, the solution is the quadric $a x^2 + b x + c = 0$,
\end{enumerate}
as is clearly expected.

Next, let us consider a more involved example.
Take the bi-variate system of two equations $\gen{ f_1 = a x^2 y^2+b x y+2=0, \ f_2 = b x+a y+2=0 }$, where  $a$ and $b$ are parameters and $x$ and $y$ are variables.
Fix the lexicographic ordering $x \succ y$.
Then, the leading terms are clearly $LT(f_1) = a x^2y^2$ and $LT(f_2) = b x$.
The Comprehensive Gr\"obner basis is as follows:
\begin{enumerate}
\item Case-1: for $a=b=0$, the solution set empty;
\item Case-2: $a\neq 0, b=0$, the solution space is given by $a y + 2=0= -2 x^2 - a$;
\item Case-3: $a=0, b\neq 0$, the solution space is given by $-y +1 = 0 = b x + 2$;
\item Case-4: $a b\neq 0$, the solution space is given be $-a^3 y^4 - 4a^2 y^3 + (b^2-4) a y^2 + 2 b^2 y - 2 b^2 = 0$ and $b x + a y + 2=0$;
\end{enumerate}
the last three cases can, of course, be checked by simple substitution.


\section{Numerical Algebraic Geometry and Cheater's Homotopy}
\label{paramotopy_sec}
Having expounded on the virtues of the CGB, we now introduce a parallel method, which attacks our problem from an entirely different perspective.
The numerical polynomial homotopy continuation (NPHC) method \cite{SW:95} is a recently introduced numerical method that finds all the solutions of the given system of polynomial equations.
It has been used in various problems in particle theory and statistical mechanics in Refs.~\cite{Mehta:2009,Mehta:2009zv,Mehta:2011xs,Mehta:2011wj,Kastner:2011zz,Nerattini:2012pi,Mehta:2012qr,Maniatis:2012ex,Mehta:2012wk,Hughes:2012hg}. Here we briefly explain the NPHC method to make the paper self-contained.

For a system of polynomial equations,
$P(x)=0$, where $P(x)=(p_{1}(x),\dots,p_{m}(x))^{T}$ and $x=(x_{1},\dots,x_{m})^{T}$, which is \textit{known to have isolated solutions}, the \textit{Classical B\'ezout Theorem} asserts that for generic values of coefficients, the maximum number of isolated solutions in $\mathbb{C}^{m}$ is $\prod_{i=1}^{m}d_{i}$, where $d_{i}$ is the degree of the $i$th polynomial.
This bound, the \emph{classical B\'ezout bound} (CBB), is
exact for generic values. The \textit{genericity} is well-defined and the interested reader is referred to Ref.~\cite{SW:95,Li:2003} for details.

Based on the CBB, a \textit{homotopy} $H(x,t)$ can be constructed as
\begin{equation}
H(x,t)=\gamma(1-t)Q(x)+t\; P(x),
\end{equation}
where $\gamma$ is a generic complex number and $Q(x)=(q_{1}(x),\dots,q_{m}(x))^{T}$ is a system of polynomial equations with the following properties:
\begin{enumerate}
\item the solutions of $Q(x)=H(x,0)=0$ are known or can be easily obtained.
$Q(x)$ is called the \textit{start system} and the solutions are
called the \textit{start solutions};
\item the number of solutions of $Q(x)=H(x,0)=0$ is equal to the CBB;
\item the solution set of $H(x,t)=0$ for $0\le t\le 1$ consists of a finite
number of smooth paths, called homotopy paths, each parameterized by
$t\in[0,1)$; and
\item every isolated solution of $H(x,1)=P(x)=0$ can be reached by some
path originating at a solution of $H(x,0)=Q(x)=0$.
\end{enumerate}
We can then track all the paths corresponding to each solution of
$Q(x)=0$ from $t=0$ to $t=1$. The paths which reach $P(x)=0=H(x,1)$ are the solutions of $P(x)=0$. By implementing
an efficient path tracker algorithm, all isolated solutions of a system of multivariate polynomials system can be obtained: it is shown \cite{SW:95} that for a generic $\gamma$, there are no singularities (i.e., paths do not cross each other) for $t\in[0,1)$. Thus, in the end, we obtain
all the solutions of the system $P(x)=0$. In this respect, the NPHC method has a great advantage over all other known methods for finding stationary points.

Several sophisticated numerical packages well-equipped with
path trackers exist, such as {\sf Bertini} \cite{BHSW06}, {\sf PHCpack}~\cite{Ver:99}, {\sf HOMPACK}~\cite{MSW:89}, and {\sf HOM4PS2}~\cite{GLW:05,Li:03},
which are all available as freewares from the respective research groups.

\subsection{Cheater's Homotopy}
The advantages of the homotopy based on the CBB are (1) the CBB is easy
to compute, and (2) the start system based on the CBB can be solved quickly.
The drawback of it is that the CBB does not take the
sparsity of the system into account; systems arising in practice
have far fewer solutions than the CBB, so a large portion
of the computational effort is wasted.

Hence, one can also use homotopies based on tighter upper bounds. For example, the 2-Homogeneous Homotopy is constructed by first writing $\mathbb{C}^m = \mathbb{C}^k\times\mathbb{C}^{k-m}$
for some $k$ where $0 < k < m$, which is accomplished by partitioning the original variables into
two groups.  This has the advantage of incorporating some of the structure of
the given polynomial system $P(x)$ into the start system $Q(x)$.
The corresponding bound, called the 2-Homogeneous B\'ezout Bound (2HomBB), is often
tighter than the CBB when the polynomial system $P(x)$ has a naturally arising
partition of the variables, which occurs in the examples below.
Given a partition, the 2HomBB is easy to compute and the start system can be
solved quickly via linear algebra.

Another important homotopy is the Polyhedral Homotopy which uses the monomial structure of the given polynomial system $P(x)$
based on the Bernstein-Khovanskii-Kushnirenko (BKK) Theorem~\cite{Bernstein75,Khovanski78,Kushnirenko76}
to yield the BKK bound. Essentially, this upper bound on the number of complex
solutions is obtained by computing the mixed volume of the convex hull
of the Newton polytope (which is based on the exponents of the monomials appearing) of each equation.
The interested reader from the physics community is referred to Ref.~\cite{Mehta:2009,Mehta:2012wk} for these above two bounds. We note that, as with the CBB, the 2HomBB and BKK bounds are also generically sharp with respect to the family of polynomial systems under consideration.

However, in the realistic systems, we do need to take the sparsity of the systems fully into account.
Indeed, even though the original system may have only a few solutions, we may have to track many paths making it computationally expensive. The {\bf cheater's homotopy} is a much more practical method to overcome this difficulty.  The crux of {\bf cheater's homotopy} relies on a theorem, which we state without its proof below, that states that the maximum number of solutions of a parametric system of polynomial equations over all the parameter-points is the one at a generic parameter-point \cite{li1989cheater}:
\begin{thm}
Let $P(x, \lambda) = 0$ be a system of polynomial equations, $p_1(x, \lambda), \dots, p_n(x,\lambda) = 0$, where $\lambda = (\lambda_1, \dots, \lambda_m) \in \mathbb{C}^m$ are parameters and $x = (x_1, \dots, x_n) \in \mathbb{C}^n$ be variables.  Then, there exists an open, dense, full-measure set $U \subset \mathbb{C}^{n+m}$ such that for\\ $(b_1^*, \dots, b_n^*, \lambda_1^*, \dots, \lambda_m^*) \in U$ the following holds:

\begin{enumerate}
\item The set $X^*$ of solutions $x = (x_1, \dots, x_n)$ of the system

$$p_1(x_1, \dots, x_n, \lambda_1^*, \dots, \lambda_m^*) + b_1^* = 0,$$
$$ \dots $$
$$p_n(x_1, \dots, x_n, \lambda_1^*, \dots, \lambda_m^*) + b_n^* = 0$$

consists of $d_0$ isolated points for $d_0 \leq d$, where $d$ is the total degree of the system for a generic $\lambda$.

\item Smoothness and accessibility properties (Properties 3 and 4 of homotopies) still hold for the cheater's homotopy which is given as follows:
$$ H(x,t) = P( x_1, \dots, x_n, (1-t)\lambda_1^* + t_1, \dots, (1-t)\lambda^*_m + t \lambda_m) + (1-t)b^* $$

where $b^* = (b_1^*, \dots, b_n^*)$.   It follows that every solution of $P(x) = 0$ is reached by a path beginning at a point of $X^*$.
\end{enumerate}

\end{thm}
Another way of viewing this is to say that a ``special" choice of our coefficients may cause the system to be deficient in the maximum number of solutions.  If we let $D$ be the set of all $\lambda$ that cause these deficiencies, then $D$ is a set of measure $0$.  Hence, cheater's homotopy relies on the fact that we choose generic, or random, values of the coefficients.  If we choose random values for $\lambda^*$, then with probability $1$, $\lambda^* \notin D$.

Algorithmically, for a given parametric system $P(x, \lambda)=0$ where $\lambda = (\lambda_1, \dots, \lambda_m)$ are parameters, we first simply need to solve $P(x, \lambda)$ at a generic parameter-point $\lambda^{*} \in \mathbb{C}^{m}$. This part has to be solved using a homotopy based on total degree, 2-homogeneous, or the BKK root counts. Then, in the second part, the system $P(x, \lambda^{*})=0$ becomes the start system of all other parameter-points $\lambda \in \mathbb{C}^{m} - \{ \lambda^{*} \}$ and the solutions of this system
become the start solutions. Finally, each path is tracked with the below homotopy:
\begin{equation}
H(x,\lambda, t)=(1-t)P(x, \lambda^{*}) + t\; P(x, \lambda).
\end{equation}
The most important trick here is to choose the generic parameter-point from the complex space. Once the generic parameter-point is chosen from the complex space, the other parameter point can be chosen to be real if the physical situation requires.  Also note that the gamma trick is implicitly employed since the $\lambda^*$ are generic.

The advantage of the cheater's homotopy over the usual NPHC is huge.  In the cheater's homotopy, the number of start solutions is drastically smaller than any of the other upper bounds usually (in the worst case, it is equal to the smallest of all the other upper bounds), and hence the number of paths to be tracked reduces a lot. A crucial advantage is that in the cheater's homotopy, we can obtain the start system for a generic parametric system once during a computationally expensive `offline' run.  This allows us to use the same start system for any number of parameter-points in the parameter-space and compute the solutions at each parameter point of interest with a much faster `online' run.

Moreover, the system at each parameter-point can be solved completely independent from any other parameter-point. In addition, while solving the system at each parameter-point, each path can also be tracked independently of all others, making it ``doubly parallelizable''. We should emphasize that the packages like {\sf Bertini} and {\sf PHCpack} now have an implementation of cheater's homotopy method in addition to the usual NPHC method.

A recently developed software module of {\sf Bertini}, called {\sf Paramotopy} \cite{paramotopytoappear}, is a different implementation of cheater's homotopy than the standard one already provided with {\sf Bertini} and can deal with a huge number of parameter points in parallel. In this paper we have extensively used this software. Several salient features of this new software are:
\begin{enumerate}
\item It offers a few options on choosing the parameter-points: the user can demand the package to discretize the parameter-space within a given range in as many parameter-points as required or the package can also be fed in a list of parameter-points provided by the user.
\item {\sf Paramotopy} first stores the data in the RAM before writing to the hard-disk which is an efficient data-management practice making the package more efficient while dealing with many parameter-points simultaneously.
\item The package is able to disregard any information other than the type of solutions the user requires in the end, i.e., it can be asked to save only real affine solutions, or only non-singular solutions, etc. This is an important aspect of the package because by not including all the information produced in a regular run, it eliminates massive data proliferation.
\item If, for any parameter-point where certain paths require higher precision, {\sf Paramotopy} informs us so that the user can re-run these specific parameter-points at higher precision settings.  Note that the parameter-points may lie on some algebraic subset in which exists an algebraic relationship on the parameters.  In this case, the number of complex solutions shrinks, and hence the related paths may require higher precision.
\end{enumerate}


\section{Illustrative Examples} \label{examples_sec}
Having explained our methods in detail, in this section, we will illustrate with examples coming from a variety of physical situations, commencing with a toy model and moving onto more serious and involved cases.

\subsection{\textit{Sys1}: A Single-Modulus Example }
We begin with a single-modulus toy example.
First, we recall that given the K\"ahler potential $K$ and superpotential $W$, both as polynomials in fields $\phi_{A=1,\dots,n}$ one can proceed to construct the scalar potential from the standard formulae \cite{Wess:1992cp}:
\begin{eqnarray}\label{potential}
V = e^K \left[ {\cal K}^{A \bar{B}} D_A W D_{\bar{B}} \bar{W} - 3
  |W|^2 \right] \ .
\end{eqnarray}
As usual the $D_A$ represents the K\"ahler derivative $\partial_A + \partial_A(K)$, and ${\cal K}^{A \bar{B}}$ is the inverse of the field space metric
\begin{equation}
{\cal K}_{A \bar{B}} = \partial_A \partial_{\bar{B}} K \ .
\end{equation}
Our example of this problem requires the solution of the critical set
\begin{equation}\label{dV}
\partial_A V = 0, \;\; \mbox{for } A = 1, \ldots, n \ .
\end{equation}
We can further classify the solutions to (\ref{dV}) by the amount of supersymmetry they preserve, the value of the bare cosmological constant they dictate,
and so forth.
The most relevant examples are:
\begin{itemize}
\item SUSY, Minkowski: $\partial_A V = 0$, $D_A W = 0$ for all $A$ and $W=0$;
\item SUSY, AdS: $\partial_A V = 0$, $D_A W = 0$ for all $A$ but $W \ne 0$.
\end{itemize}

Thus prepared, let us take a single field example, which is addressed in the demo of {\sf StringVacua} and with which we can compare.
Let the K\"ahler potential $K$ and superpotential $W$ of an ${\cal N}=1$ supersymmetric theory in a single complex moduli field $T$ be
given as
\begin{equation}
K=-3\log(T+\bar{T}) \ ,
\quad
W=a+bT^{8} \ .
\end{equation}
Note that the field $T$ comes along with its complex conjugate. Even though they can be treated as different variables by merely relabeling them, they are not actually independent variables. To avoid this problem, we can write them in
terms of real and imaginary parts, i.e., $T=t+i\,\tau$ where $t$ and $\tau$ are real.
The potential, using \eqref{potential}, is
\begin{equation}
V =  \frac{1}{3t}(4b(5b(t^{2}+\tau^{2})^{7}-3a(t^{6}-21t^{4}\tau^{2}+35t^{2}\tau^{4}-7\tau^{6})))
\end{equation}
which has $2$ variables. To find the stationary points of $V$,
we need to compute the zero locus of the partial
derivatives of $V$ with respect to variables $t$ and $\tau$:
\begin{eqnarray}
\nn
\frac{\partial V}{\partial t} & = & \frac{1}{3t^{2}}(4b(5b(13t^{2}-\tau^{2})(t^{2}+\tau^{2})^{6}-3a(5t^{6}-63t^{4}\tau^{2}+35t^{2}\tau^{4}+7\tau^{6})))=0,\\
\frac{\partial V}{\partial\tau} & = & \frac{1}{3t}(56b\tau(5b(t^{2}+\tau^{2})^{6}+a(9t^{4}-30t^{2}\tau^{2}+9\tau^{4})))=0 \ .
\end{eqnarray}
For general values of parameters $a$ and $b$ the system already becomes difficult to analyze using symbolic methods and one could solve the system for specific values such as $a=b=1$ \cite{Gray:2006gn,Gray:2009fy}.

We also note that the stationary equations in this example involve
denominators. Since we are not interested in the solutions for which
the denominators are zero, we clear them out by multiplying them with
the numerators appropriately. In these equations, all the denominators
are multiples of $t$. The condition that none of the denominators
is zero can be imposed algebraically by adding an additional equation,
$1-y\, t=0$, with $y$ being an additional variable. Thus, we now have $3$ equations in $3$ variables.
The CGB library of {\sf Singular} can deal with this system.
The expression of the CGB of this system is quite large so we do not write it down here.

\begin{figure}[t!]
\centering
\includegraphics[width= 0.59\linewidth]{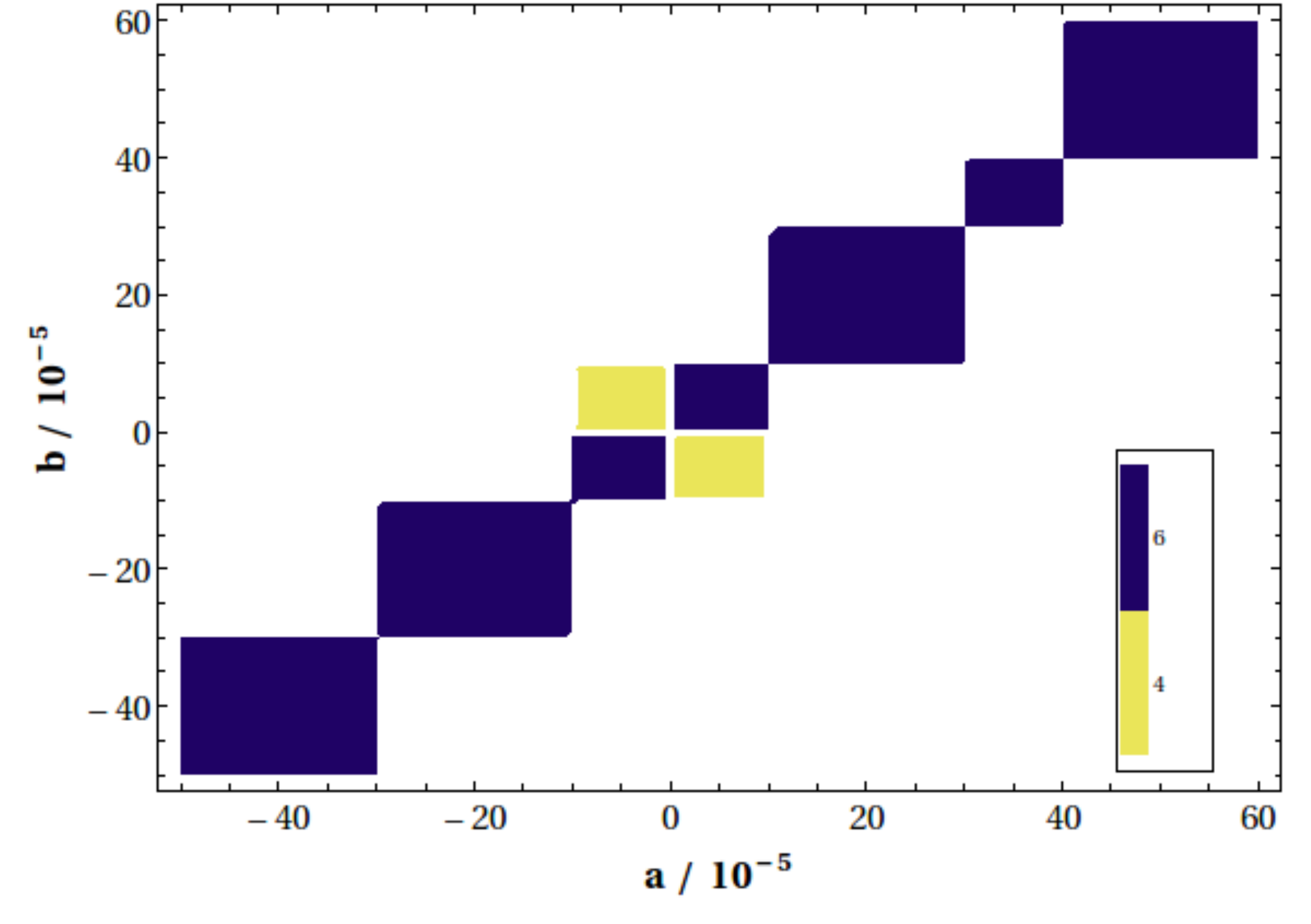}
\caption{Scanned values for $a$ and $b$ for the one-modulus example in \textit{Sys1}.
The corresponding number of solutions of the system per parameter point is indicated by the color. The total number of parameter points is 100,150 and the spacing between the points is equidistant.}
\label{sys1_numsol_fig}
\end{figure}

We can also solve this system using Cheater's homotopy. For the range of values of $a$ and $b$ given in Figure~\ref{sys1_numsol_fig} we find that there are either $6$ solutions for $a,b>0$ and $a,b<0$ or $4$ solutions for $a<0,\,b>0$ and $a>0,\,b<0$; these are indicated by different colors in Figure~\ref{sys1_numsol_fig}.
We scan over a total of 100,150 parameter points and find a total of 582,676 solutions. Exactly half of the solutions are physical, i.e. $t>0$ which corresponds to a positive volume of the cycle associated to the K\"ahler modulus $T$. We did our computation on a desktop machine on single processor (Linux machine with 2.1GHz cloak speed). First, Bertini takes around $30$ minutes to solve the system from scratch for a given parameter-point. This means that it would have taken around $5.731$ years to solve the system at all the $100,150$ parameter-points. With the CH method, however, we solved them in only $55$ hours.

\begin{figure}[h!!!]
\centering
\includegraphics[width= 0.49\linewidth]{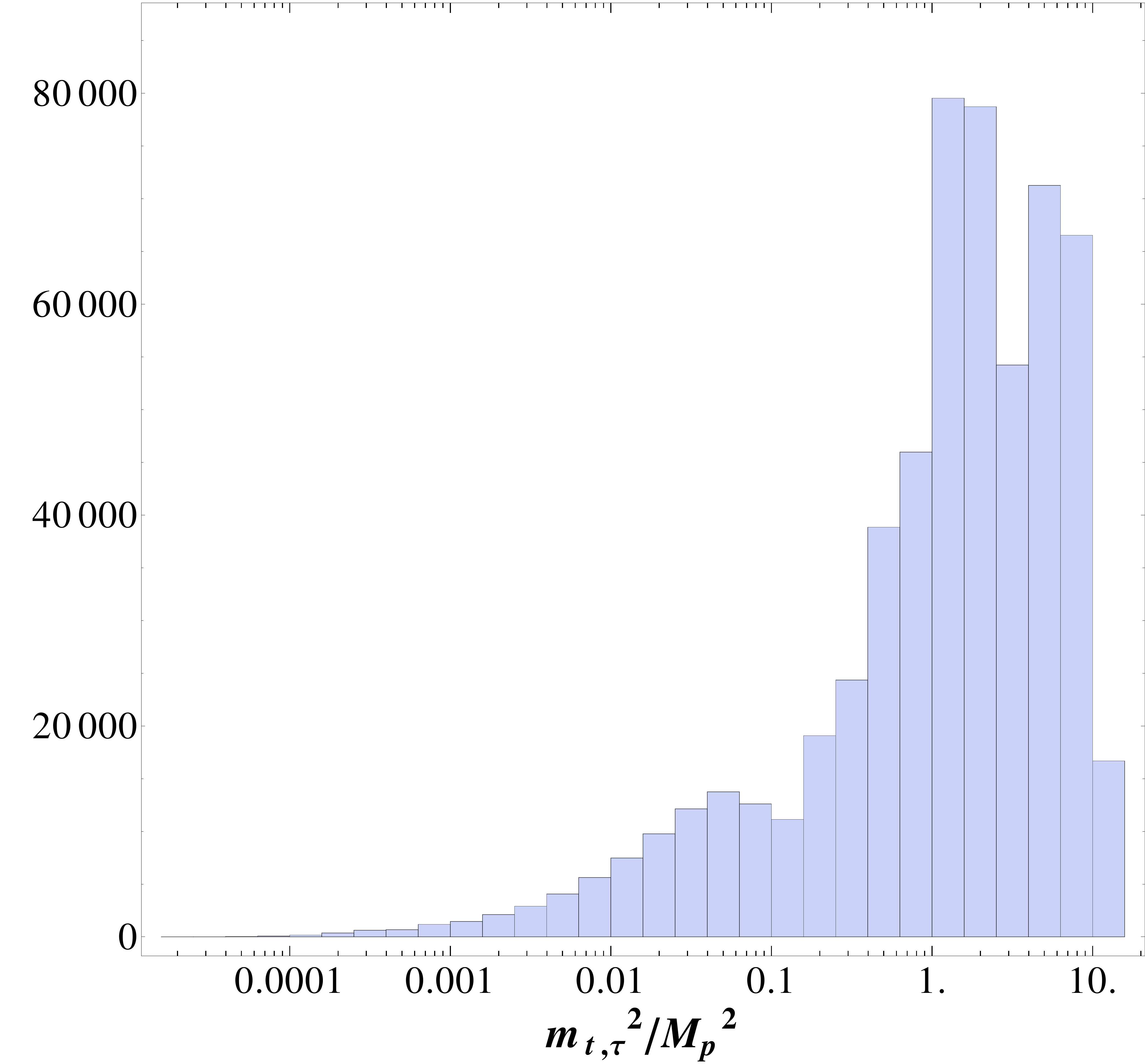}
\includegraphics[width= 0.49\linewidth]{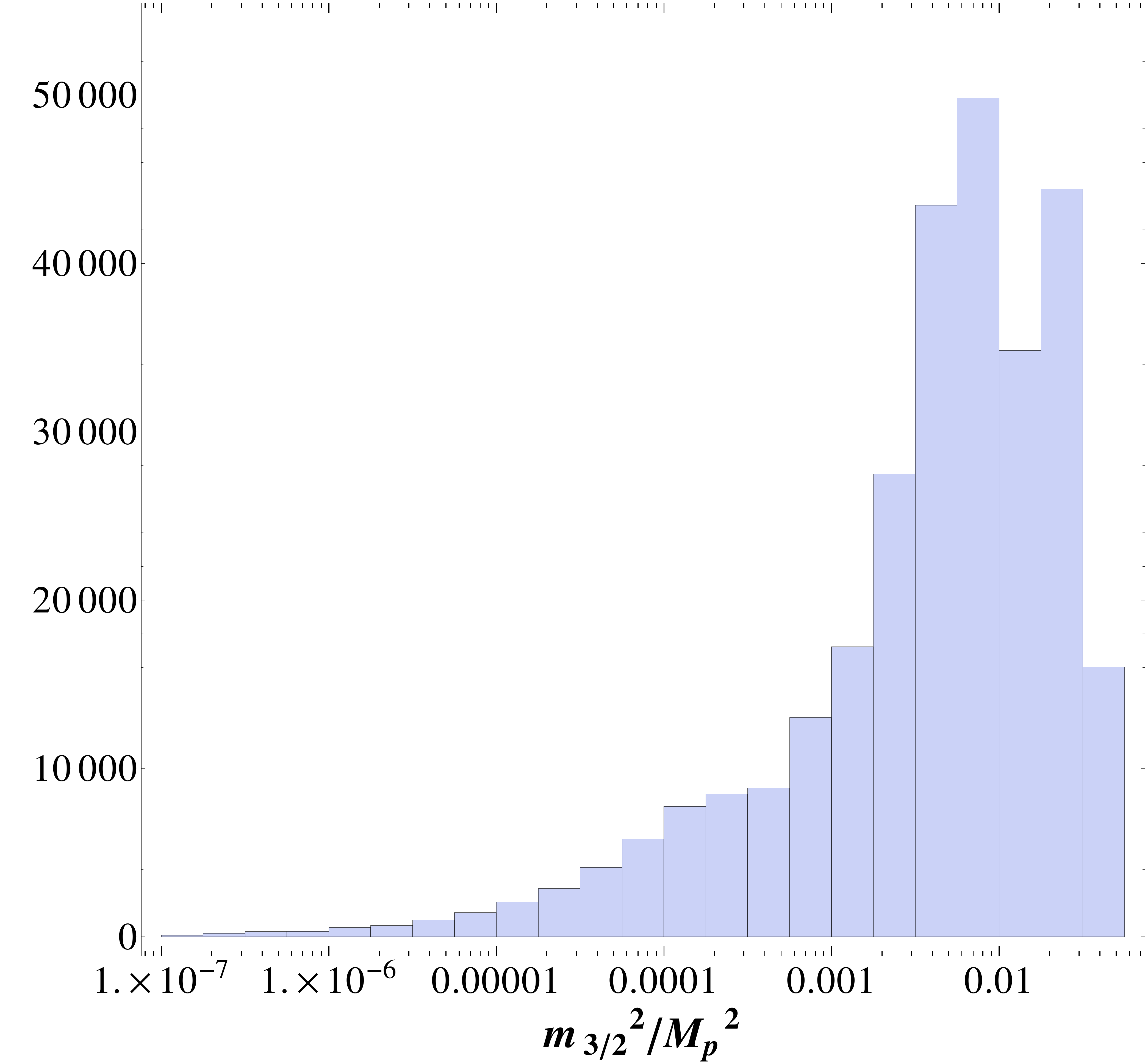}
\caption{Masses of the moduli $t$ and $\tau$ as eigenvalues of the Hessian (left) and the gravitino mass (right) for the one-modulus model \textit{Sys1}.}
\label{sys1_masses_fig}
\end{figure}

The scalar masses of the moduli $t$ and $\tau$ which can be calculated as the eigenvalues of the Hessian of $V$ are found to be positive for all parameter points with $t>0$. We give the moduli masses and the gravitino mass
\begin{equation}
 m_{3/2}^2 = e^K\, |W|^2\,,
\end{equation}
which determines the scale of supersymmetry breaking in Figure~\ref{sys1_masses_fig}, which is a frequency plot of the mass values for our space of $100,150$ parameters.

\subsection{\textit{Sys2}: A Two-Moduli Model}
Now, let us move on to a model with two moduli fields.
Consider the K\"ahler potential and superpotential
\begin{eqnarray}
\nn K & = & -3\log(T+\bar{T})-\log(S+\bar{S}),\\
W & = & aS+bST+cT^{2},
\end{eqnarray}
with two fields, a K\"ahler modulus $T=t+i\tau$ and the axio-dilaton $S=s+i\sigma$.
Hence, using \eqref{potential}, the stationary equations to be solved are:
\begin{eqnarray}
\nn
0 &=& (3a^2(s-\sigma)(s+\sigma)+6at(b(-s^2+\sigma^2)+2c\sigma\tau)+2bc\sigma\tau(t^2-3\tau^2)\\
\nn
& &-b^2(s-\sigma)(s+\sigma)(5t^2-3\tau^2) +c^2(5t^2-3\tau^2)(t^2+\tau^2)), \\
\nn
0 &=& (-9a^2(s^2+\sigma^2)+b^2(s^2+\sigma^2)(5t^2-9\tau^2)+2bc\tau(\sigma t^2+18st\tau-9\sigma\tau^2)\\
\nn
& & +c^2(-5t^4+2t^2\tau^2-9\tau^4)+6a(2b(s^2+\sigma^2)t+c(5st^2+4\sigma t\tau-3s\tau^2))), \\
\nn
0 &=& (-(c\sigma t(6a+bt))+(6acs+3b^2(s^2+\sigma^2)-18bcst-2c^2t^2)\tau+9bc\sigma\tau^2+6c^2\tau^3),\\
\nn
0 &=& (3a^2 \sigma-6at(b\sigma+c\tau) +b(-5b\sigma t^2-ct^2\tau+3b\sigma\tau^2+3c\tau^3)),\\
\nn
0 &=& 1-zts \ .
\end{eqnarray}
The first four equations arise from setting the numerators of the various partial derivatives of $V$ to zero, and the last is an auxiliary equation to ensure that the denominator does not vanish.
Therefore, this system is $5$ equations in $5$ variables with $3$ parameters.
Again, this is an example used in {\sf StringVacua}, but $a,b$ and $c$ are parameters that were chosen to be $1,-1,1$ respectively in~\cite{Gray:2006gn,Gray:2009fy}.
Now,  we can undertake the much more challenging task of taking a huge range of parameter-values in our computation.

\begin{figure}[h!!!]
\centering
\includegraphics[width= 0.59\linewidth]{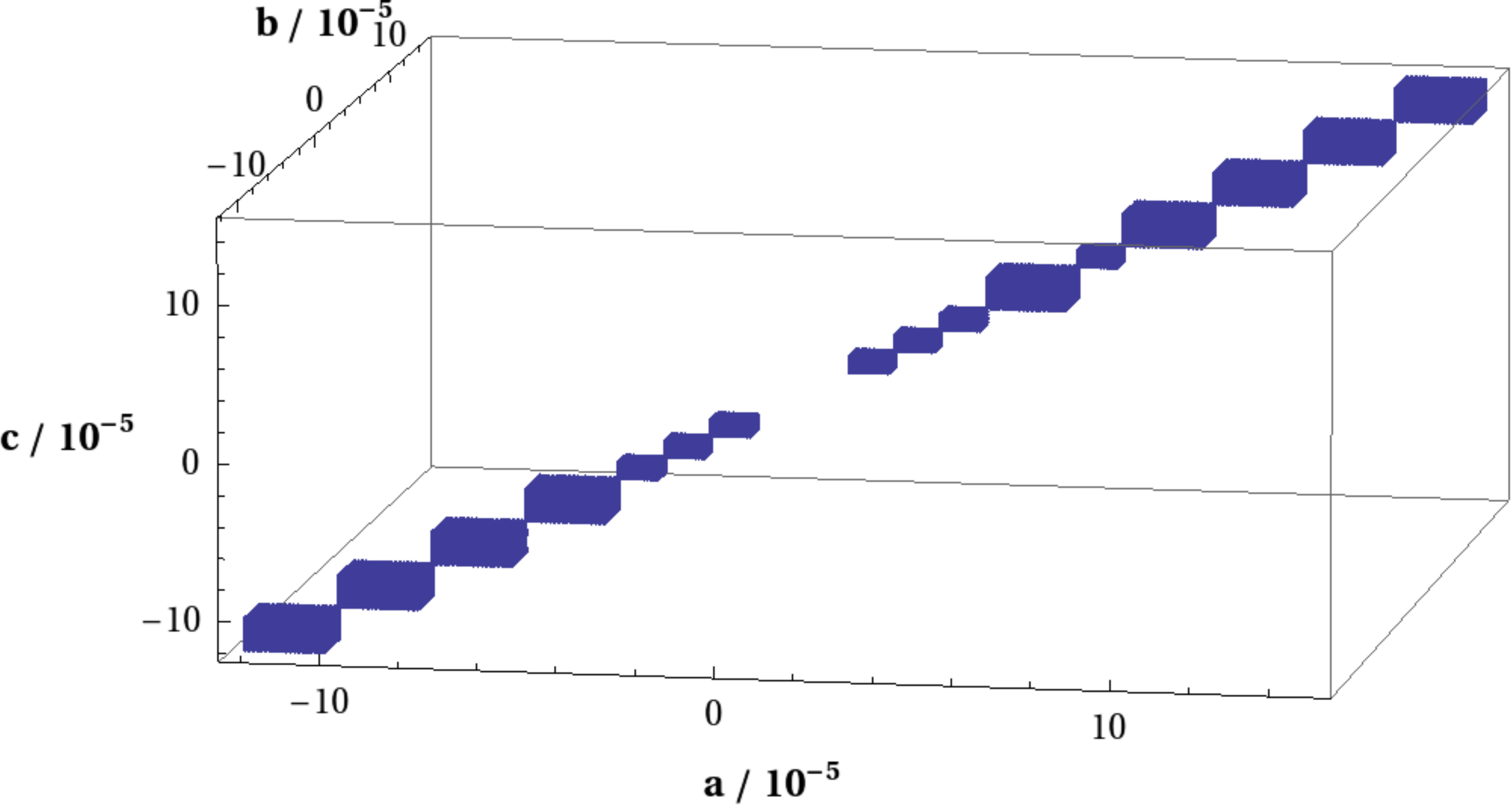}
\caption{Scanned values for $a$ and $b$ and $c$ for the two-moduli example \textit{Sys2}. The total number of parameter points is $100,672$, and the spacing between the points is equidistant.
The blue regions indicate that there are 6 solutions to \textit{Sys2}.}
\label{sys2_points_fig}
\end{figure}

Again, for generic choice of the values for $a,b$ and $c$, it becomes difficult for the traditional GB method. However, we can compute the CGB for this system; The output from {\sf Singular} of the CGB for this system is quite large, and, hence, we do not write it here.

Using {\sf Paramotopy} on the other hand, we easily scan over $100,672$ parameter points, see Figure~\ref{sys2_points_fig}.
We find $6$ solutions per parameter point which yields a total of $604,032$ solutions. The physicality condition for the case of this model demands again $t>0$ and also $s=g_s^{-1}>0$, i.e. positive string coupling. This is fulfilled by $503,299$ solutions. As for \textit{Sys1}, we did our computation on a desktop machine. {\sf Bertini} takes around $40$ minutes to solve the system from scratch for a given parameter-point. Hence, for all the $100,672$ parameter-points, it would have taken $7.682$ years. However, using the CH method, we solved them all in $125$ hours. When evaluating the Hessian of $V$, we find that there is always at least one negative eigenvalue, i.e. no vacua exist for this model for the set of parameters we considered.

\subsection{\textit{Sys3}: The Quintic}
An area of string phenomenology where one can make use of the power of Cheater's homotopy is the landscape of flux vacua in type IIB string theory~\cite{Giddings:2001yu,Douglas:2006es,Grana:2005jc,Blumenhagen:2006ci}.
As mentioned in \textit{Sys1}, if one goes to a particular corner of the moduli space, the equations
\begin{equation}
 D_i W =0 \label{susyvac}
\end{equation}
that determine the supersymmetric vacuum state of the no-scale scalar potential~\cite{Cremmer:1983bf,Ellis:1983sf}
\begin{equation}
 V = e^{K} K^{a\bar{b}} D_a W \overline{D_b W}\,,\label{V4Dnoscale}
\end{equation}
are polynomial equations in the complex structure moduli fields $\phi_a = \tau,U_1,..,U_{h^{2,1}}$, where $\tau=\sigma + i\, s \,(= i\bar S)$ is the axio-dilaton. The parameters of these equations are the flux integers
\begin{align}
 \begin{aligned}
  &\frac{1}{(2\pi)^2 \alpha'} \int_{A_a} F_3 = {f_1}_a \in \mathbb{Z}\,,\qquad \frac{1}{(2\pi)^2 \alpha'} \int_{B^a} F_3 = {f_2}_a \in \mathbb{Z}\,,\\
 &\frac{1}{(2\pi)^2 \alpha'} \int_{A_a} H_3 = {h_1}_a \in \mathbb{Z}\,,\qquad \frac{1}{(2\pi)^2 \alpha'} \int_{B^a} H_3 = {h_2}_a \in \mathbb{Z}\,,
 \end{aligned}
\end{align}
where $F_3$ and $H_3$ are the RR and NS three-form flux of type IIB string theory, and $\gen{A_a,B^b}$ is a symplectic basis for the $b_3 = 2 h^{2,1} + 2$ three-cycles. The flux integers are to be chosen freely as long as the D3 tadpole constraint
\begin{equation}
 L = \frac{1}{(2\pi)^4 (\alpha')^2} \int_{X_3} H_3 \wedge F_3 =  h_1f_2-h_2f_1 \label{D3tad}
\end{equation}
is not violated.

For supersymmetric flux configurations, $F_3$ and $H_3$ always combine into an imaginary-self-dual (ISD) flux $G_3 = F_3 - \tau \,H_3$~\cite{Giddings:2001yu,Denef:2004ze}, which can be written as~\cite{Ibanez:2012zz}
\begin{equation}
 \ast_6 s\,H_3 = -(F_3 - \sigma H_3)\,.\label{ISDF3H3}
\end{equation}
Consequently, one only has to consider $2h^{2,1}+2$ independent directions of the original $4h^{2,1}+4$ flux integers defining
\begin{equation}
 H_3 = \left( \begin{array}{c}
         h_1\\ h_2
        \end{array}
 \right) \qquad \quad \text{and} \qquad \quad F_3 = \left( \begin{array}{c}
         -h_2\\ h_1
        \end{array}
 \right)\,,\label{h1h2}
\end{equation}
Then, the D3 tadpole eq.~\eqref{D3tad} manifests as a positive definite form, i.e.
\begin{equation}
 L = h_1^2 + h_2^2\,,
\end{equation}
for a symplectic basis of three-cycles. With this form of the D3 tadpole, we can use {\sf Paramotopy} to find all flux vacua, i.e. solutions to eq.~\eqref{susyvac} for all flux configurations $h_1$, $h_2$ with D3 tadpole $L<L_{max}$ for a given maximal tadpole $L_{max}$. We have to take into account the $SL(2,\mathbb{Z})$ invariance of IIB string theory in order to consider only physically equivalent flux configurations. It was shown in~\cite{MartinezPedrera:2012rs} that for fluxes of the form~\eqref{h1h2}, only the configurations
\begin{equation}
 \left( \begin{array}{c}
         h_1\\ h_2
        \end{array}
 \right) \cong
 \left( \begin{array}{c}
         -h_1\\ -h_2
        \end{array}
 \right) \cong
 \left( \begin{array}{c}
         -h_2\\ h_1
        \end{array}
 \right) \cong
 \left( \begin{array}{c}
         h_2\\ -h_1
        \end{array}
 \right)\,.\label{4h1h2}
\end{equation}
are related by $SL(2,\mathbb{Z})$, and hence physically equivalent.

This problem to find all supersymmetric flux vacua in the large complex structure limit was carried out for the compactification manifold $\mathbb{CP}^4_{11169}[18]$ in~\cite{MartinezPedrera:2012rs} in the context of de Sitter model building in the K\"ahler uplifting scenario~\cite{Rummel:2011cd,Louis:2012nb}.  To present a simple example, we consider the one complex structure modulus $\psi$ of the mirror quintic which has $h^{1,1}=101$ and $h^{2,1}=1$. In the limit of large complex structure, this describes the one moduli subspace of the quintic~\cite{Candelas:1990rm} given by the vanishing of the polynomial
\begin{equation}
 x_0^5+x_1^5+x_2^5+x_3^5+x_4^5-5\,\psi\,x_0x_1x_2x_3x_4\,,
\end{equation}
in $\mathbb{CP}^4_{11111}$ with coordinates $x_0,..,x_4$. To obtain a polynomial system let us introduce the complex coordinate $U=\nu + i\, u$ which relates to $\psi$ as
\begin{equation}
 U\simeq -\frac{5}{2\pi i} \log (5\psi)\,,\label{Upsi}
\end{equation}
up to corrections of $\mathcal{O}(U^k e^{2\pi i U})$ that are exponentially suppressed in the large complex structure limit $\text{Im}(U)\to \infty$. This moduli space can be described by an approximately polynomial prepotential in the large complex structure limit
\begin{equation}
 \mathcal{G} = \omega_0^2\left(-\frac{5}{6} \left(\frac{\omega_1}{\omega_0}\right)^3 -\frac{11}{4} \left(\frac{\omega_1}{\omega_0}\right)^2 +\frac{25}{12}\frac{\omega_1}{\omega_0} - \frac{25\zeta(3)\chi}{2 (2\pi i)^3}   \right)\,,\label{prepot}
\end{equation}
with $\chi = 2(h^{1,1}-h^{2,1}) = 200$ and $\omega_a$ and $\mathcal{G}_b$ are the periods
\begin{equation}
 \omega_a = \int_{A_a} \Omega\,,\qquad\qquad \mathcal{G}_b = \int_{B^b} \Omega\,.\label{defperiods}
\end{equation}
for $a,b=1,2$ and $\mathcal{G}_b = \partial_{\omega_b} \mathcal{G}$. $\omega_0$ can be interpreted as the normalization $\Omega$ of the holomorphic three-form, such that the one physical variable is $U=\omega_1 / \omega_0$. Eq.~\eqref{prepot} is valid up to corrections $\mathcal{O}(e^{2\pi i k U})$ that are exponentially small in the limit of the large complex structure. We define the large complex structure limit as $\text{Im}(U) > 2$ such that these corrections to $\mathcal{G}$ are smaller than $10^{-3} \mathcal{G}$.\footnote{Note that there is a conifold singularity at $\psi^5 =1$ that is excluded from the large complex structure limit $\text{Im}(U) > 2$ according to eq.~\eqref{Upsi}.}

Together with the axio-dilaton, $\tau$ this yields a two moduli example described by the K\"ahler potential and superpotential
\begin{align}
\begin{aligned}
K &= K_{\text{k}} -\log \left(-i \int_{X_3} \Omega(U) \wedge \bar{\Omega}(\bar{U}) \right) - \log \left(-i(\tau-\bar \tau) \right)\,,\\
&= -\log \left(i \sum_{a=1}^2 (\bar{\omega}_a \mathcal{G}_a - \omega_a \bar{\mathcal{G}}_a) \right) - \log \left(-i(\tau-\bar \tau) \right)\,,\\[2mm]
 W_0 &= \frac{1}{2\pi}\, \int_{X_3} (F_3 - \tau H_3) \wedge \Omega(U)\,,\\
  &=2\pi \sum_{a=1}^2 \left[({f_1}_a-\tau\,{h_1}_a)\mathcal{G}_a- ({f_2}_a-\tau\,{h_2}_a)U_a\right]\,,\label{KW0period}
\end{aligned}
\end{align}
where $K_{\text{k}}$ is the K\"ahler potential of the K\"ahler moduli and setting $\alpha'=1$.

\begin{figure}[h!!!]
\centering
\includegraphics[width= 0.320\linewidth]{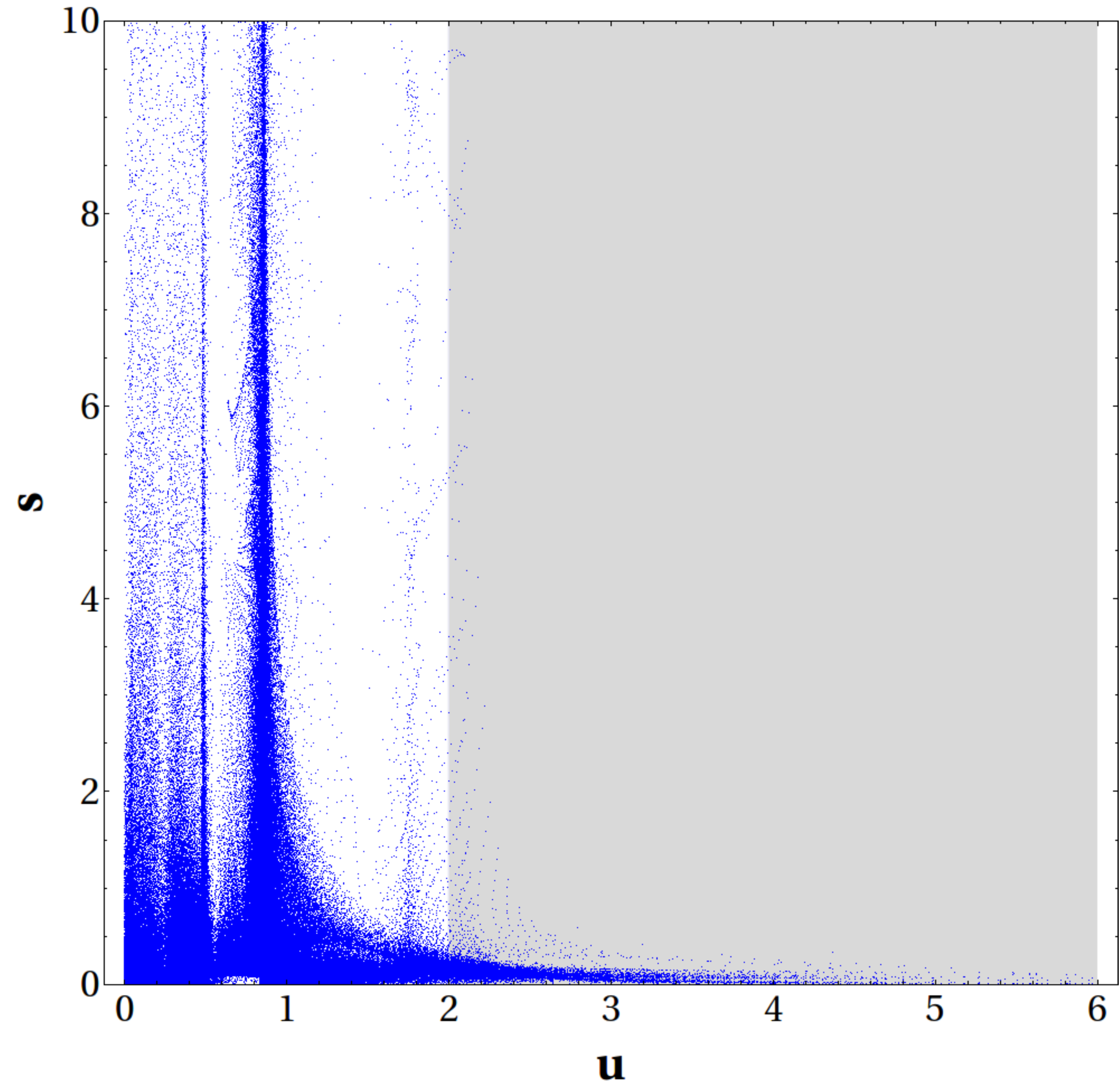}
\includegraphics[width= 0.330\linewidth]{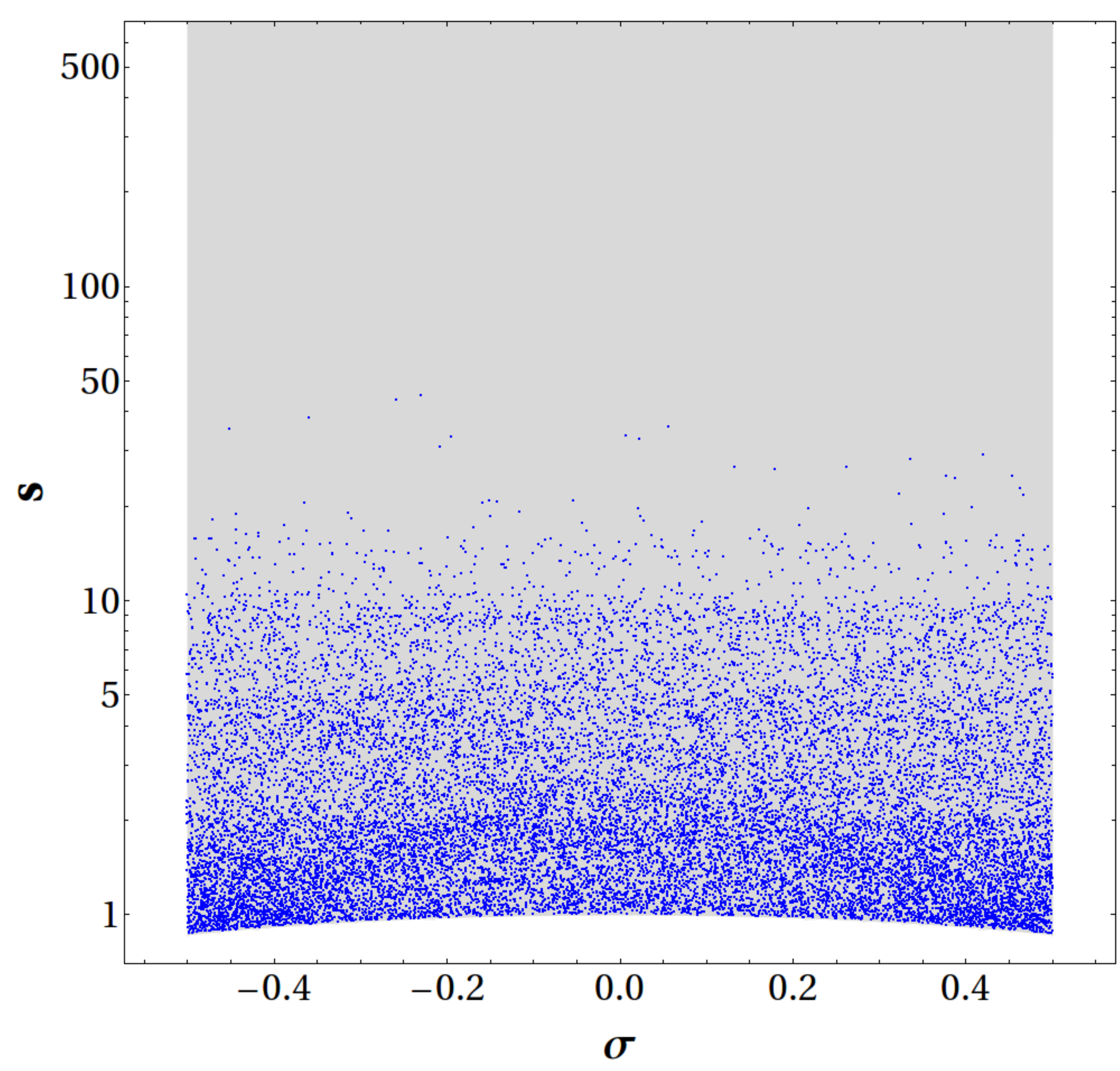}
\includegraphics[width= 0.333\linewidth]{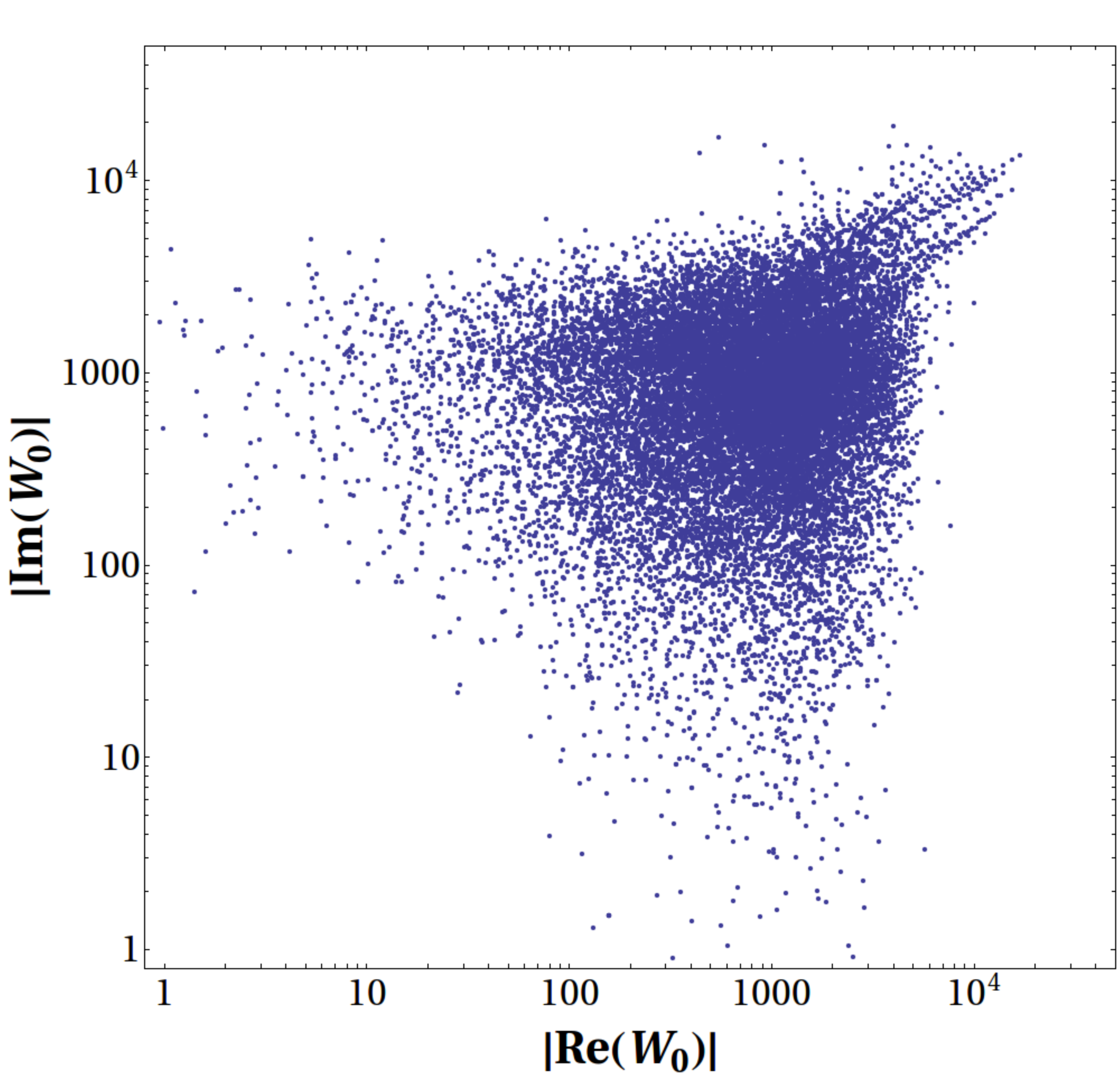}
\caption{Left: Distribution of \textit{all} algebraic solutions to eq.~\eqref{susyvac} with $s>0$ and $u>0$ for the K\"ahler potential and superpotential \eqref{KW0period} in \textit{Sys3}. The gray shaded area indicates the space of physical solutions, i.e. $s>0$ and $u>2$. Middle: Distribution of the dilaton $\tau=\sigma + i\, s$. Right: Distribution of the superpotential $W_0$.}
\label{quintic_LCStauW0_fig}
\end{figure}

We solve eqs.~\eqref{susyvac} for \eqref{KW0period} up to a maximal tadpole of $L_{max}=625$. This corresponds to $481,825$ flux parameter points for which we find a total of $1,726,334$ solutions, i.e. on average $\sim 3.6$ solutions per parameter point. Only $20,280$ are physical solutions, i.e. $g_s^{-1} = s>0$ and $u>2$, see Figure~\ref{quintic_LCStauW0_fig}. If we were to solve it at each parameter-point from scratch, then it takes 72 minutes per parameter-point. This means that for total $481,825$ parameter-points that we solved the system for, we would have taken $481,825 \times 72/60 = 578,190$ hours. But instead, using {\sf Paramotopy}, we solved all of them in only $3776$ hours, i.e., around $26.4$ human hours with a cluster of $144$ processors.  We could not obtain the CGB for this system in a reasonable time.

We can make use of the $SL(2,\mathbb{Z})$ symmetry of IIB string theory, to transform each solution to the fundamental domain
\begin{equation}
 -\frac{1}{2} \leq \text{Re}(\tau) \leq \frac{1}{2} \qquad \text{ and } \qquad |\tau|>1\,,\label{funddomain}
\end{equation}
via the successive transformations
\begin{equation}
 \tau' = \tau+b\,, \quad W_0' = W_0\,, \qquad \text{and} \qquad \tau' = -1/\tau\,, \quad W_0' = W_0/\tau \label{S+btrans}
\end{equation}
with $b\in \mathbb{Z}$. We show the distribution of the obtained values for $\tau=\sigma + i\, s$ and the flux superpotential $W_0$ in Figure~\ref{quintic_LCStauW0_fig}. We see that the the strongly coupled region $s = 1/g_s \sim 1$ is preferred and values of $W_0 \sim \mathcal{O}(10^2-10^3)$ are preferred. The same qualitative behavior was observed in~\cite{MartinezPedrera:2012rs} for the manifold $\mathbb{CP}^4_{11169}[18]$.

\begin{figure}[h!!!]
\centering
\includegraphics[width= 0.49\linewidth]{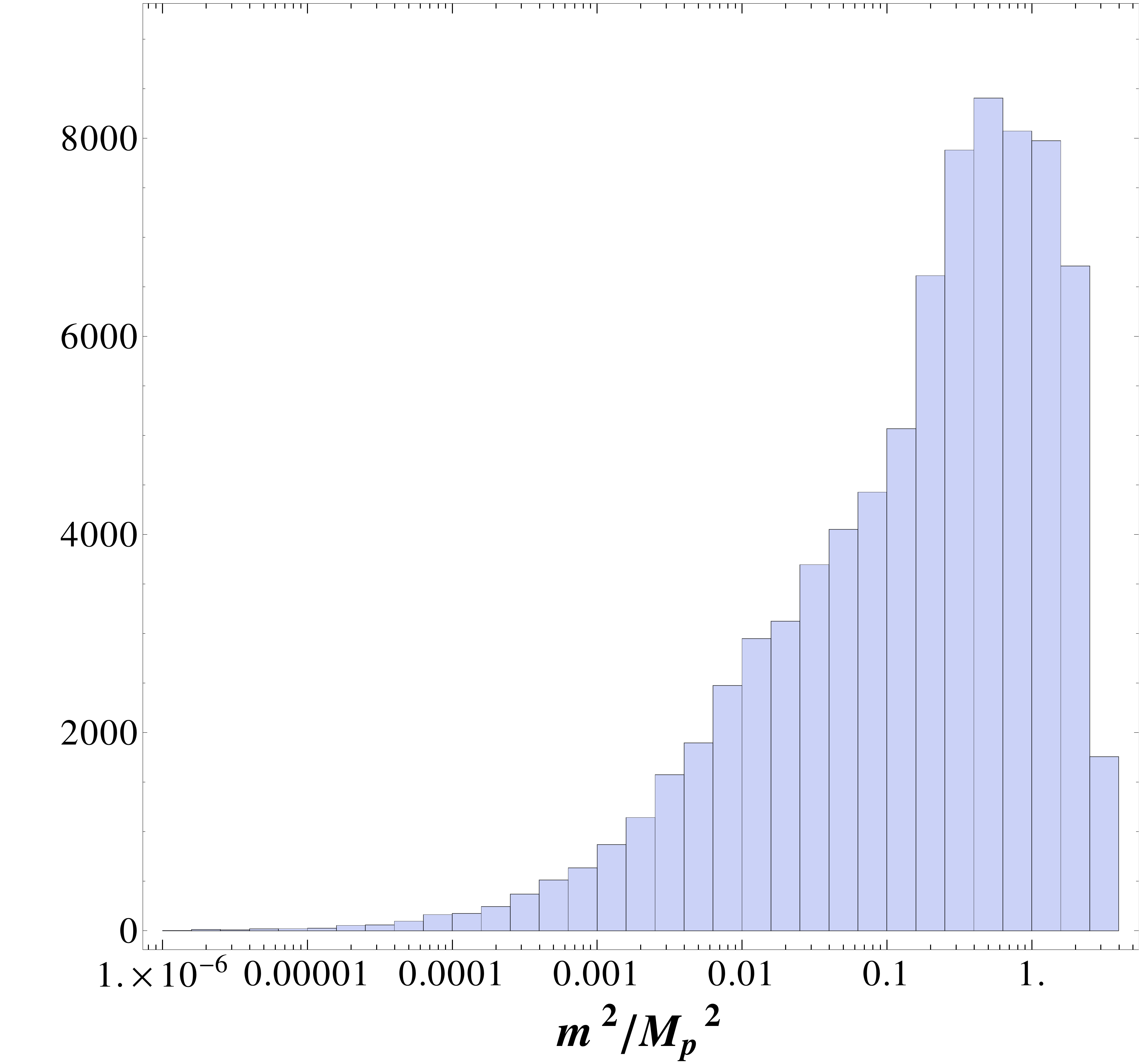}
\includegraphics[width= 0.49\linewidth]{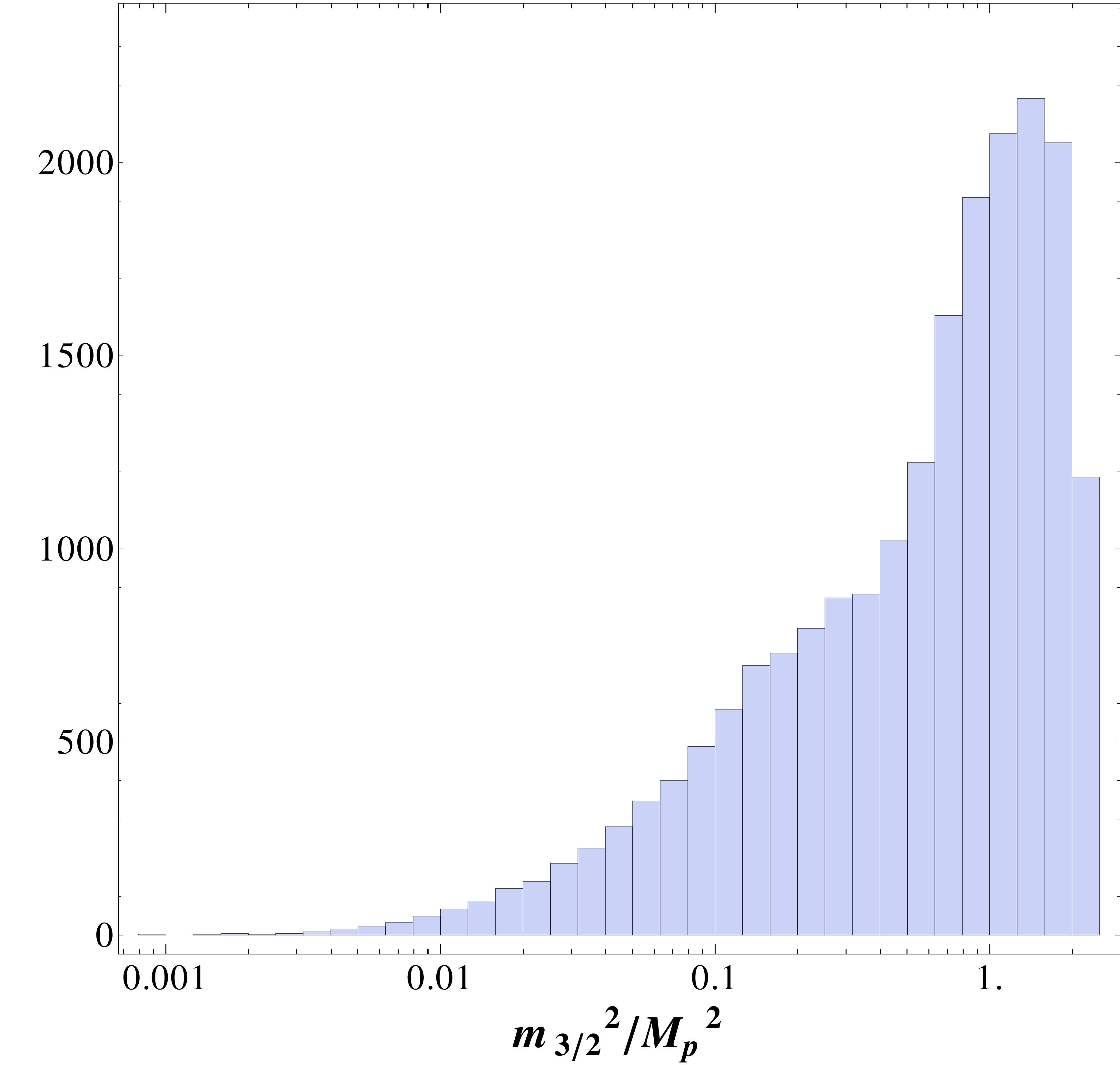}
\caption{For \textit{Sys3} (\eqref{susyvac} and \eqref{KW0period}), the distribution of the masses $m^2$ of the moduli $t,\tau,s$ and $\sigma$, i.e. all eigenvalues of the Hessian of $V$ (left) and the gravitino mass $m_{3/2}^2$ (right) in units of $M_{\text{P}}^2$ for $\mathcal{V}=100$.}
\label{quintic_massmod_fig}
\end{figure}

Finally, we give the masses $m^2$ of the moduli $t,\tau,s$ and $\sigma$ and the gravitino mass $m_{3/2}^2$ in Figure~\ref{quintic_massmod_fig}. For this, we have to specify the value of the K\"ahler moduli K\"ahler potential in eq.~\eqref{KW0period}. For a Calabi-Yau compactification, this is given as $K_{\text{k}}=-2\,\log \mathcal{V}$, where $\mathcal{V}$ is the volume modulus of the Calabi-Yau and its vacuum expectation value depends on the stabilization mechanism of the K\"ahler moduli. Here we choose $\mathcal{V}=100$ in string units for definiteness.

\subsection{Further Applications: \textit{Sys4}}
As mentioned in the introduction, large systems with a multitude of parameters are ubiquitous.
Thus, our methods above should be applicable to far more situations than finding the extrema of a PEL in the context of effective supersymmetric Lagrangians.
In this subsection, as a parting example, let us see the power of Cheater's homotopy applied to the geometry of Calabi-Yau manifolds.

For concreteness and continuing along the vein of \textit{Sys3}, let us focus on the problem of finding the singular locus (or the absence thereof) given a family of quintics.\footnote{
Recently there has been much activity in study the database of so-called complete intersection Calabi-Yau (CICY) manifolds of which the quintic is the simplest example as well their smooth quotients.
Checking smoothness for models with this database of 7890 Calabi-Yau threefolds and their descendents, for example, is a crucial step \cite{Candelas:1987kf,Anderson:2007nc,Braun:2010vc}.
}

Let us consider the family of quintic manifolds given by as a homogeneous hypersurface in $\mathbb{C}\mathbb{P}^4$ with coordinates $x_0,\dots,x_4$, and let
\begin{equation}
Q(x_0, \dots, x_4; a,b,c) =
x_0^5 + x_1^5 +x_2^5 +x_3^5 + x_4^5 + a x_0 x_1 x_2 x_3 x_4 + b x_0^4 x_1 + c x_1^3 x_2 x_3
,\label{Qabc}
\end{equation}
where $a,b,c$ are complex parameters.
Of course, the most general family of quintics has 101 deformation parameters corresponding to $h^{2,1}$ of the manifold, but this example suffices to show the power of the method.
Now, for $b=c=0$, we have the one most familiar to us, with $a = -5$ being the conifold point in the complex structure moduli space.
The singular locus of $Q$ is given by the Jacobian ideal $\gen{\partial_{x_i} Q}$ for $ i = 0, \dots, 4 $, excluding the origin and defined up to scaling of the projective coordinates.

\begin{figure}[t!!!]
\centering
\includegraphics[width=0.48\linewidth]{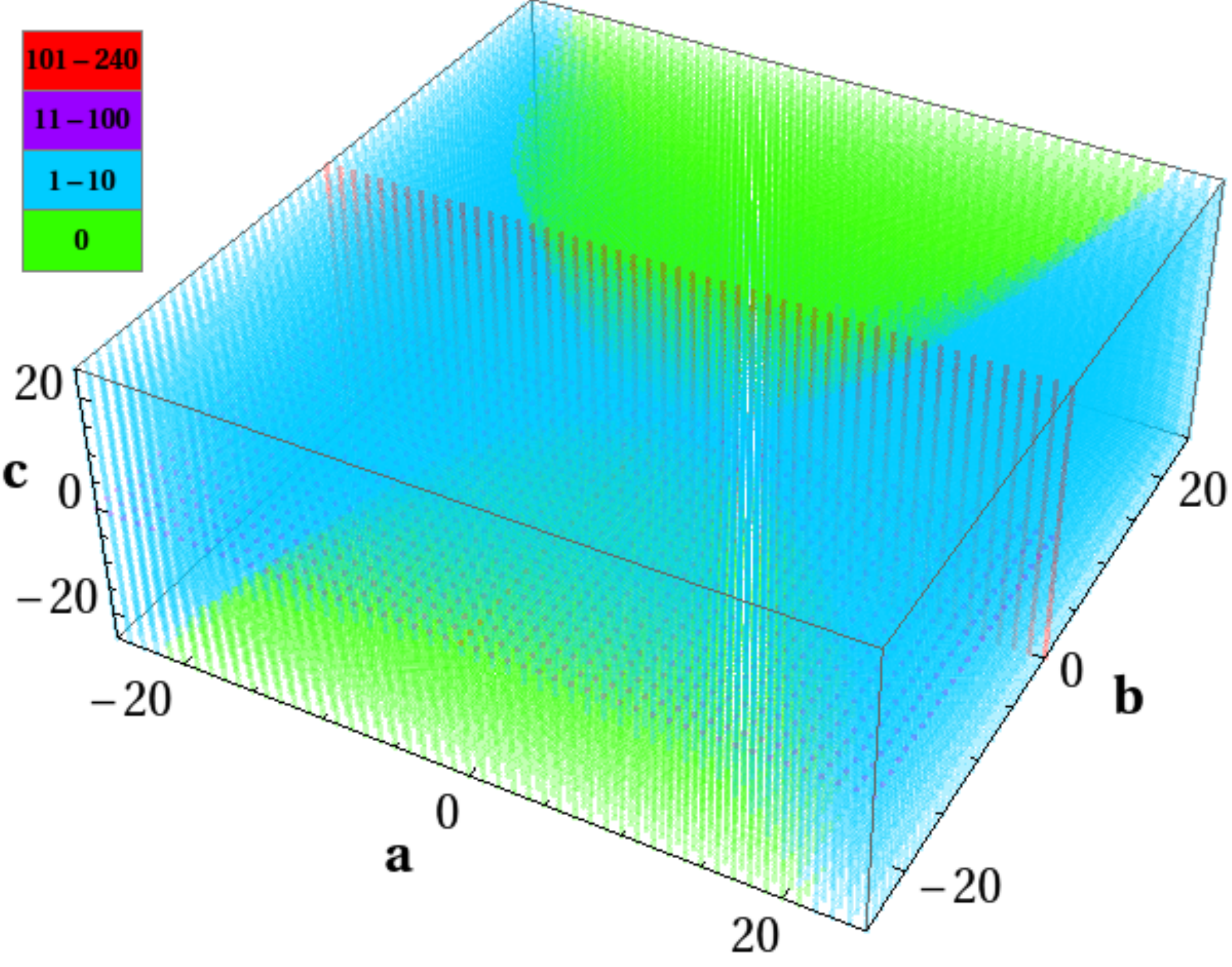}
\hskip 3mm
\includegraphics[width=0.48\linewidth]{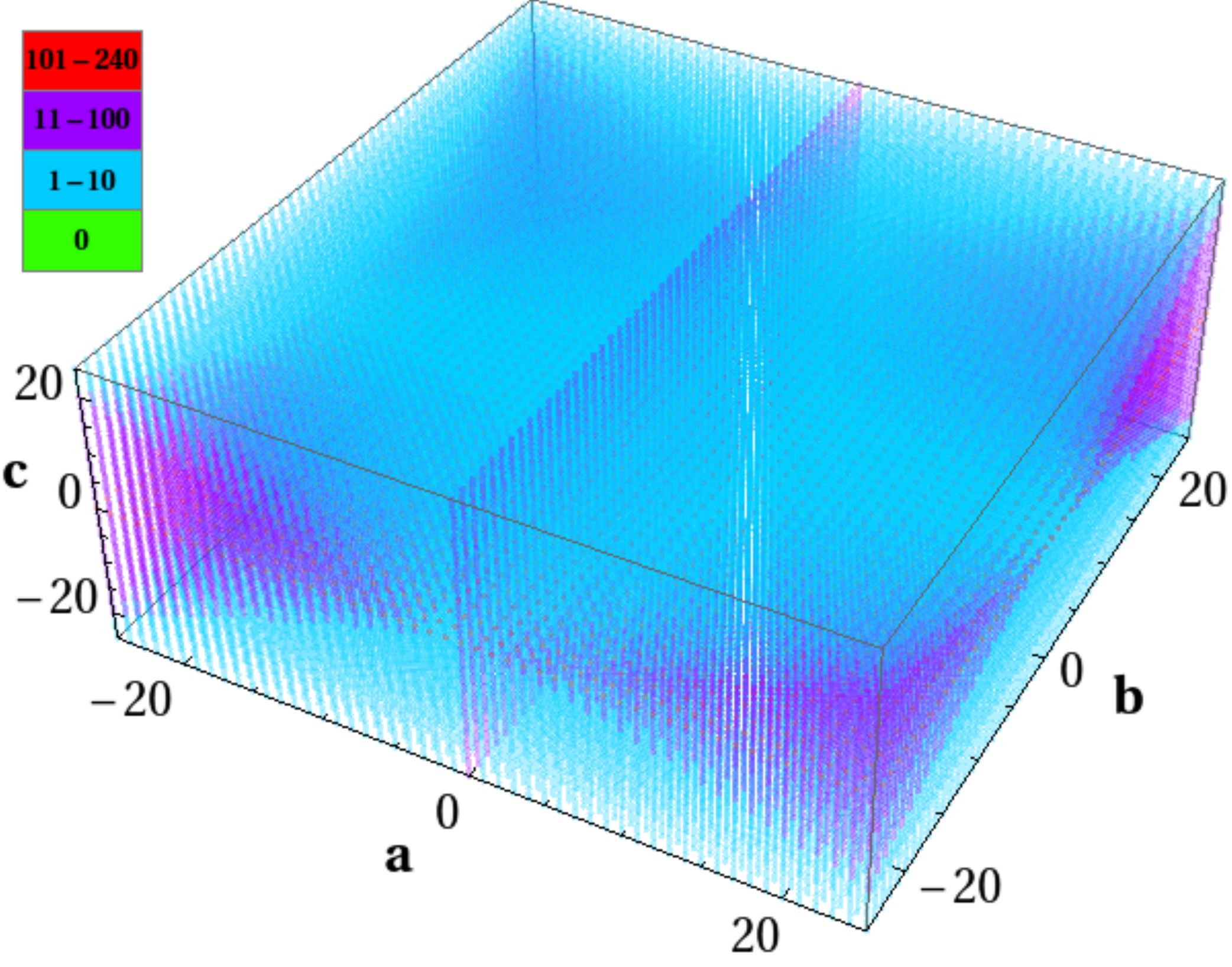}\\
\includegraphics[width=0.48\linewidth]{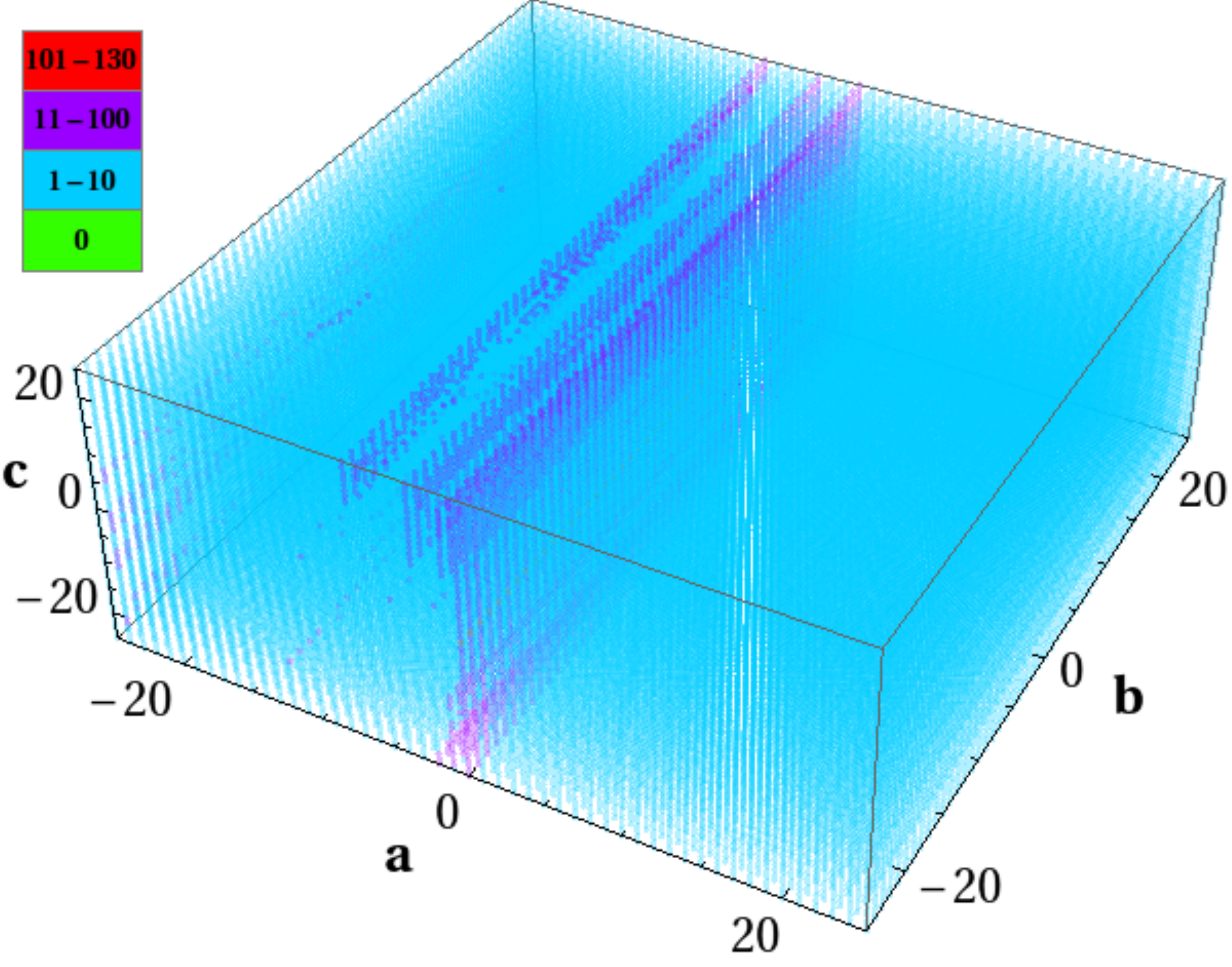}
\hskip 3mm
\includegraphics[width=0.48\linewidth]{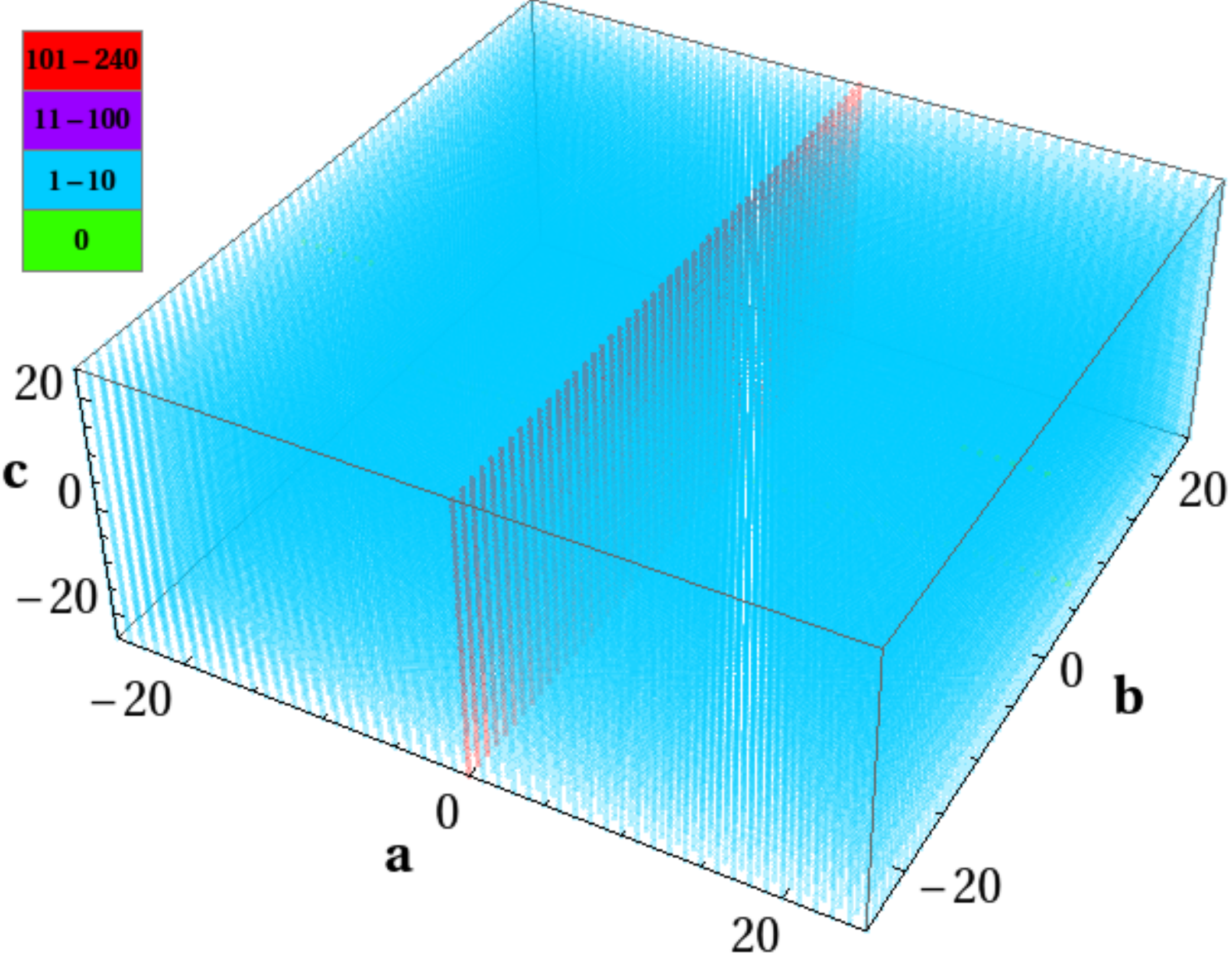}
\caption{The number of solutions at each parameter point $\{a,b,c\}$ for system $P_0$ (top left), $P_1$ (top right), $P_2$ and $P_3$ (bottom left) and $P_4$ (bottom right) indicated by the different colors.}
\label{quinticsing_nsol_fig}
\end{figure}

It is more convenient to work over affine patches of $\mathbb{C}\mathbb{P}^4$ which can be rescaled to be $P_i = \{ x_i = 1\}$ for $ i = 0, \dots, 4 $.
For each of the five patches, we compute the four remaining partial derivatives, the solution of which is then the system we need to analyze.
That is, we need to perform {\sf Paramotopy} on
\begin{align}
\begin{aligned}
\partial_{x_{j \ne i}} Q(x_i = 1; a,b,c) = 0 \ , \quad i = 0, \dots, 4 \ .
\end{aligned}
\end{align}
All of the 5 systems are of the same size and we take $a,b,c \in [-25, 25]$ in increments of one, i.e. $51^3 = 132,651$ parameter points. {\sf Bertini} takes from 144 to 223 seconds at a parameter-point if solved from scratch, hence it would have taken from 0.6057 to 0.938854 years for all 132651 parameter-points. {\sf Paramotopy} solved each of these systems for all parameter-points in from 19.216 to 92.9002 hours (using 72 processors, it took 960.812 to 4645.01 seconds). Obtaining the CGB for this system is quite fast as well; however, obtaining the solutions for each parameter-point from the CGB takes the same amount of time as {\sf Paramotopy}.  Hence, we show the results only obtained from {\sf Paramotopy} here. We show the number of solutions per parameter point indicated by the different colors in Figure~\ref{quinticsing_nsol_fig} and Table~\ref{quinticsingnsol_tab}. The solution space of the systems $P_2$ and $P_3$ is identical since eq.~\eqref{Qabc} is invariant under $x_2 \leftrightarrow x_3$.

\begin{table}[h!]
\centering
  \begin{tabular}{|c||c|c|c|c|}
  \hline
  $\# sol$ & $0$ & $1-10$ & $11-100$ & $101 - 240$\\
  \hline
  \hline
  $P_0$ & $35,613$ & $93,162$ & $1,250$ & $2,626$\\
  \hline
  $P_1$ & $0$ & $109,326$ & $20,724$ & $2,601$\\
  \hline
  $P_2,P_3$ & $0$ & $126,159$ & $6,441$ & $51$\\
  \hline
  $P_4$ & $60$ & $129,990$ & $0$ & $2,601$\\
  \hline
  \end{tabular}
  \caption{Distributions of the number of solutions $\# sol$ for the five systems $P_i$.}
  \label{quinticsingnsol_tab}
\end{table}

\section{Conclusion and Outlook}\label{conc_sec}
Parametric systems of non-linear equations arise very naturally and frequently in theoretical physics.
The parameters add one more hurdle in solving such systems in addition to the non-linearity of the equations as one has to solve the system many times.
In the present paper, we have demonstrated that two of the sophisticated algebraic geometry methods can solve such systems very efficiently provided that the nonlinearity is polynomial-like.

The first method is the Comprehensive Gr\"obner basis (CGB) which is a symbolic method based on Gr\"obner bases. Given a (non-parametric) system of equations, we can systematically find another system of equations which has the same solution space as the original one and is easier to solve, called a Gr\"obner basis.
If we have a parametric system of equations, then we can compute the CGB which is a Gr\"obner basis for all values of the parameters including their special values.
To obtain the solutions of the system at a specific parameter-point, one then needs to input the values of the parameters in the CGB and solve the system using the traditional methods.
However, the algorithmic complexity to compute the CGB is the same, or in many cases even worse, as the Gr\"obner basis method, i.e., the memory required by the algorithm blows up exponentially with increasing number of variables, equations, monomials, and degree.
Moreover, the related algorithm is highly sequential and unparallelizable.
Hence, though the CGB method is extremely useful in the cases where we can finish the computation, but for many physical systems the method often falls short.

The second method we described in this paper is called the cheater's homotopy.
Here, one first solves the given parametric system of polynomial equations at a generic parameter-point because it is shown that the number of solutions at such a parameter-point is an upper bound on the number of solutions at any other parameter-points.
Then, one tracks solution-paths from the solutions at the generic parameter-point to the ones at which the system needs to be solved using the numerical polynomial homotopy continuation method.
This method is highly parallelizable, and, hence, we can solve the system at thousands of parameter-points in a short time on a computer cluster.
We use a package called {\sf Paramotopy} which is not only a highly efficient implementation of the related algorithm, but also does data and memory management effectively. These features allow for a huge number of parameter-points to be solved simultaneously.

We have used these two methods on various examples arising from computing the extrema of potential energy landscapes, especially in the context of moduli stabilization in string phenomenology. However, obtaining the solutions for each parameter-point from the CGB (for the cases it is possible to obtain the CGB) takes the same amount of time as {\sf Paramotopy}; hence, we show the results only obtained from the cheater's homotopy in this paper.

The examples are given as a 4D $\mathcal{N}=1$ effective supergravity scalar potential, defined by a K\"ahler potential $K$ and a superpotential $W$.
Our first two examples \textit{Sys1} and \textit{Sys2} are toy models of one and two complex scalar fields, respectively.

We looked for stationary points of these scalar potentials for $\sim 10^5$ parameter points, finding half of the stationary points are vacua in the case of \textit{Sys1} and finding no vacua in the case of \textit{Sys2}, i.e., at least one tachyonic direction always exists.
Finally, we looked at flux compactifications on a realistic Calabi-Yau, the quintic in the case of two moduli. We look for flux vacua for $\sim 5\cdot 10^5$ flux configurations, which corresponds to a maximal D3 tadpole of $L=625$.
Our results indicate preferences of strongly coupled vacua $g_s \gtrsim 1$ and values of the superpotential $W_0 \sim \mathcal{O}(10^2-10^3)$.

The applicability of the methods described here is, of course, much wider and is intended for any large parametric polynomial systems, with the parameters appearing explicitly as coefficients.
Indeed, these methods should provide an extremely powerful tool for many areas in theoretical physics such as potential energy landscape, statistical mechanics, particle phenomenology, string phenomenology, and non-linear dynamics.
As an illustration, we have applied Cheater's homotopy to the geometry of Calabi-Yau manifolds in the case of \textit{Sys4}.

If we were to solve \textit{Sys1}, \textit{Sys2}, the quintic, and the Calabi-Yau systems from scratch for each parameter-point using the numerical homotopy continuation method, a single processor machine would have taken $5.731$ years, $7.682$ years, $66.003$ years, and 0.6057-0.938854 years, respectively. But using the cheater's homotopy method we solved them in 55 hours, 125 hours, 3776 hours and 19.216-92.9002 hours, respectively. This is a drastic reduction of the computational efforts, see also Table~\ref{comptimeconcl_tab}.

\begin{table}[h!]
\centering
  \begin{tabular}{|c||c|c|c|}
  \hline
  System & Number of parameter points & single-processor& Cheater's homotopy\\
  \hline
  \hline
  \textit{Sys1} & $100,150$ & $5.731$ yr & $55$ h\\
  \hline
  \textit{Sys2} & $100,672$ & $7.682$ yr & $125$ h\\
  \hline
  \textit{Sys3} & $481,825$ & $66.003$ yr & $3776$ h\\
  \hline
  \textit{Sys4} & $132,651$ & $0.6-0.9$ yr & $19.2-92.9$ h\\
  \hline
  \end{tabular}
  \caption{Number of parameter points and computation time for the examples under consideration.}
  \label{comptimeconcl_tab}
\end{table}

We emphasize that the two methods are based on complex algebraic geometry which means that the variables are first considered to be complex, and then only purely real solutions are retained, if desired.
Because in many applications we only have real parameters and only real solutions are physically important, a natural question to ask is if any method exists which can deal with systems with both parameters and variables defined over reals.
Indeed, a more recent method exists from symbolic real algebraic geometry, based on the discriminant varieties~\cite{hanan2010stability,hernandez2011towards}, which treats both variables and parameters as reals from the beginning and gives a different set of information than the methods described in this paper.
This certainly constitutes a direction worthy of pursuit.
On the numerical side, little progress exists in this direction, and so far methods exist to extract only limited information out for the whole parameter space.
For example, in Ref.~\cite{griffin2012real}, a numerical method is proposed which can tell us the maximum and minimum number of real solutions over the whole parameter-space. It is our hope that the applications in this paper will also become a motivation to develop such methods and making them more for real-life applications.

\section*{Acknowledgements}

DM would like to thank the U.S. Department of Energy
for their support under contract no. DE-FG02-85ER40237; to Tianran Chen and Tim McCoy for their helpful remarks on this work; and to the entire Bertini, PHCpack and HOM4PS2 groups for their continuous support in this project. A major part of this paper is based on DM's PhD thesis. MN wishes to thank the NSF for their partial support under grant DMS-1025564. YHH would like to thank the Science and
Technology Facilities Council, UK, for an Advanced Fellowship and grant ST/J00037X/1,
the Chinese Ministry of Education, for a Chang-Jiang Chair Professorship at NanKai University,
the U.S. National Science Foundation for grant CCF-1048082, as well as City University,
London and Merton College, Oxford, for their enduring support. MR would like to thank Alexander Westphal for helpful discussions.

\comment{

The Comprehensive Gr\"obner basis computation from {\sf Singular} for \textit{Sys1} yields:
\begin{enumerate}
\item For $b=a=t y-1 = 0$, the whole $\mathbb{C}^3$ is the solution;
\item For $b=0=a$, the solution space is given by the solutions of the equations:
\begin{eqnarray*}
      0 & = & tau^12*y^12*b^2+6*tau^10*y^10*b^2+15*tau^8*y^8*b^2+20*tau^6*y^6*b^2+15*tau^4*y^4*b^2+6*tau^2*y^2*b^2+b^2,\\
      0 & = & t*b^2+tau^12*y^11*b^2+6*tau^10*y^9*b^2+15*tau^8*y^7*b^2+20*tau^6*y^5*b^2+15*tau^4*y^3*b^2+6*tau^2*y*b^2, \\
      0 & = & t*y-1,
\end{eqnarray*}
\
[3]:

   [1]:

      1

   [2]:

      _[1]=b

      _[2]=t*y-1

[4]:

   [1]:

      a*b

   [2]:

      _[1]=81*y^32*a^5*b+8620209*y^24*a^4*b^2-1558051056*y^16*a^3*b^3-8770320384*y^8*a^2*b^4+66560000000*a*b^5

      _[2]=27*tau*y^24*a^4*b+2873520*tau*y^16*a^3*b^2-506898432*tau*y^8*a^2*b^3-5120000000*tau*a*b^4

      _[3]=10355329932897064960000000*tau^2*a*b^5+912933335703789*y^30*a^5*b+97142317113223534221*y^22*a^4*b^2-19070230452836330366064*y^14*a^3*b^3+80813474166329917768704*y^6*a^2*b^4

      _[4]=10355329932897064960*tau^2*y^2*a*b^4-14178650157*y^24*a^4*b-1509776776899405*y^16*a^3*b^2+179661842602824432*y^8*a^2*b^3-750183244746224640*a*b^4

      _[5]=2704987661344560*tau^2*y^8*a^2*b^3+42010815280000000*tau^2*a*b^4-280011627*y^22*a^4*b-30435810687675*y^14*a^3*b^2+131893770976032*y^6*a^2*b^3

      _[6]=225057939*tau^2*y^10*a^2*b^2-13519740864*tau^2*y^2*a*b^3-51543*y^16*a^3*b-284232903*y^8*a^2*b^2+1232643776*a*b^3

      _[7]=31112403*tau^2*y^16*a^3*b-70702220208*tau^2*y^8*a^2*b^2-1069040000000*tau^2*a*b^3+735200544*y^14*a^3*b-3185869024*y^6*a^2*b^2

      _[8]=78421391360000000*tau^3*a*b^4-137334879*tau*y^22*a^4*b-14622078389040*tau*y^14*a^3*b^2+1923043935704064*tau*y^6*a^2*b^3

      _[9]=15684278272*tau^3*y^2*a*b^3-1197*tau*y^16*a^3*b-131056656*tau*y^8*a^2*b^2-5208552448*tau*a*b^3

      _[10]=244150896*tau^3*y^8*a^2*b^2-10640000000*tau^3*a*b^3-56223*tau*y^14*a^3*b-341992800*tau*y^6*a^2*b^2

      _[11]=399*tau^3*y^10*a^2*b-1221312*tau^3*y^2*a*b^2+9501*tau*y^8*a^2*b+399808*tau*a*b^2

      _[12]=1069040000000*tau^4*a*b^3-10413627*tau^2*y^14*a^3*b+14546503872*tau^2*y^6*a^2*b^2-244150896*y^12*a^3*b+1057987216*y^4*a^2*b^2

      _[13]=22908*tau^4*y^4*a*b^2-15*tau^2*y^10*a^2*b-18504*tau^2*y^2*a*b^2-228*y^8*a^2*b+988*a*b^2

      _[14]=399808*tau^5*y^2*a*b^2-279*tau^3*y^8*a^2*b-270272*tau^3*a*b^2-4389*tau*y^6*a^2*b

      _[15]=12669*tau^5*y^8*a^2*b-20000000*tau^5*a*b^2+288570*tau^3*y^6*a^2*b-206151*tau*y^4*a^2*b

      _[16]=20000000*tau^7*a*b^2-22521*tau^5*y^6*a^2*b-414630*tau^3*y^4*a^2*b+139359*tau*y^2*a^2*b

      _[17]=tau^7*y^6*a*b+21*tau^5*y^4*a*b-49*tau^3*y^2*a*b+11*tau*a*b

      _[18]=65*tau^12*y^12*b^2+390*tau^10*y^10*b^2+975*tau^8*y^8*b^2-12*tau^6*y^14*a*b+1300*tau^6*y^6*b^2-135*tau^4*y^12*a*b+975*tau^4*y^4*b^2+198*tau^2*y^10*a*                                                                                                                                                             b+390*tau^2*y^2*b^2-15*y^8*a*b+65*b^2

      _[19]=65*t*b^2+65*tau^12*y^11*b^2+390*tau^10*y^9*b^2+975*tau^8*y^7*b^2-12*tau^6*y^13*a*b+1300*tau^6*y^5*b^2-135*tau^4*y^11*a*b+975*tau^4*y^3*b^2+198*tau^2                                                                                                                                                             *y^9*a*b+390*tau^2*y*b^2-15*y^7*a*b

      _[20]=t*y-1

      _[21]=11*t*tau*a*b+tau^7*y^5*a*b+21*tau^5*y^3*a*b-49*tau^3*y*a*b

)
}

\bibliographystyle{JHEP}
\bibliography{bibliography_NPHC_NAG}


\end{document}